\documentclass[aps,prl,twocolumn,showpacs,superscriptaddress]{revtex4-1}

\usepackage{graphicx}
\usepackage{dcolumn}
\usepackage{color}
\usepackage{array,multirow,makecell}
\usepackage{bm}
\usepackage{nicefrac}

\begin{document}
\title{Low-dimensional antiferromagnetic fluctuations in the heavy-fermion paramagnetic ladder UTe$_2$}

\author{W. Knafo}
\affiliation{LNCMI-EMFL, CNRS UPR3228, Univ. Grenoble Alpes, Univ. Toulouse, Univ. Toulouse 3, INSA-T, 143 Avenue de Rangueil, 31400 Toulouse, France}

\author{G. Knebel}
\affiliation{Univ. Grenoble Alpes, CEA, Grenoble INP, IRIG, PHELIQS, 38000, Grenoble, France}

\author{P. Steffens}
\affiliation{Institut Laue Langevin, 6 rue Jules Horowitz, BP 156, 38042 Grenoble, France}

\author{K. Kaneko}
\affiliation{Materials Sciences Research Center, Japan Atomic Energy Agency, Tokai, Ibaraki 319-1195, Japan}

\author{A. Rosuel}
\affiliation{Univ. Grenoble Alpes, CEA, Grenoble INP, IRIG, PHELIQS, 38000, Grenoble, France}

\author{J.-P. Brison}
\affiliation{Univ. Grenoble Alpes, CEA, Grenoble INP, IRIG, PHELIQS, 38000, Grenoble, France}

\author{J. Flouquet}
\affiliation{Univ. Grenoble Alpes, CEA, Grenoble INP, IRIG, PHELIQS, 38000, Grenoble, France}

\author{D. Aoki}
\affiliation{Institute for Materials Research, Tohoku University, Ibaraki 311-1313, Japan}

\author{G. Lapertot}
\affiliation{Univ. Grenoble Alpes, CEA, Grenoble INP, IRIG, PHELIQS, 38000, Grenoble, France}

\author{S. Raymond}
\affiliation{Univ. Grenoble Alpes, CEA, IRIG, MEM, MDN, 38000 Grenoble, France}


\date{\today}

\begin{abstract}
Inelastic-neutron-scattering measurements were performed on a single crystal of the heavy-fermion paramagnet UTe$_2$ above its superconducting temperature. We confirm the presence of antiferromagnetic fluctuations with the incommensurate wavevector $\mathbf{k}_1=(0,0.57,0)$. A quasielastic signal is found, whose momentum-transfer dependence is compatible with fluctuations of magnetic moments $\mu\parallel\mathbf{a}$, with a sine-wave modulation of wavevector $\mathbf{k}_1$ and in-phase moments on the nearest U atoms. Low dimensionality of the magnetic fluctuations, consequence of the ladder structure, is indicated by weak correlations along the direction $\mathbf{c}$. These fluctuations saturate below the temperature $T_1^*\simeq15$~K, in possible relation with anomalies observed in thermodynamic, electrical-transport and nuclear-magnetic-resonance measurements. The absence or weakness of ferromagnetic fluctuations, in our data collected at temperatures down to 2.1 K and energy transfers from 0.6 to 7.5 meV, is emphasized. These results constitute constraints for models of magnetically-mediated superconductivity in UTe$_2$.

\end{abstract}

\maketitle

\begin{figure}[t]
\includegraphics[width=1\columnwidth,trim={3.2cm 1.8cm 2.4cm 1.6cm},clip]{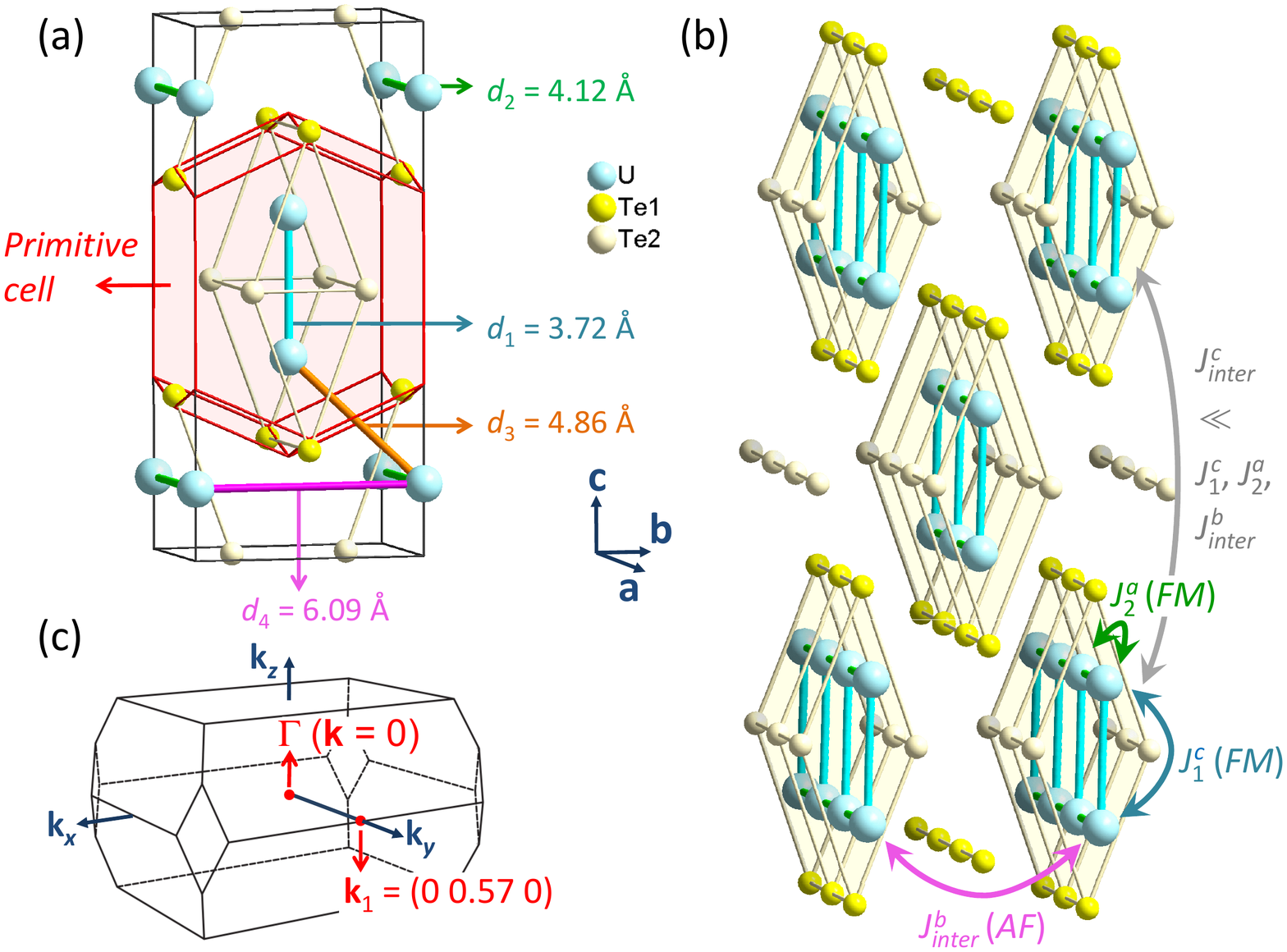}
\caption{(a) Orthorhombic and Wigner-Seitz primitive cells, and identification of the four smallest distances (from neutron diffraction at $T=2.7$~K \cite{Hutanu2020}), (b) extended structure emphasizing the ladder structure of U atoms, and (c) Brillouin zone of UTe$_2$. A phenomenological magnetic-exchange scheme implied in the magnetic-fluctuations mode with wavevector $\mathbf{k}_1$ is indicated in (b) (details are given in the text).}
\label{Fig1}
\end{figure}

The discovery of superconductivity at temperatures below $T_{sc}\simeq1.6$~K in the heavy-fermion paramagnet UTe$_2$ \cite{Ran2019,Aoki2019b} opened a breach in correlated-electron physics. UTe$_2$ was presented as a candidate for topological superconductivity, whose triplet and chiral characters have been proposed \cite{Ishizuka2019,Ran2019,Aoki2019b,Nakamine2019,Metz2019,Kittaka2020,Jiao2020,Shishidou2021,Hayes2021,Bae2021,Sato2017}. It turned rapidly out that this system is  a unique model to study the electronic correlations and their feedback on magnetism and superconductivity \cite{Shick2021,Ishizuka2021,Miao2020,Xu2019,Machida2020}. A competition between different superconducting pairing mechanisms was indicated from the observation of multiple superconducting phases stabilized near quantum magnetic instabilities under pressure and magnetic field \cite{Braithwaite2019,Knafo2019,Miyake2019,Knebel2019,Ran2019b,Aoki2020,Ran2020,Lin2020,Knebel2020,Knafo2021,Aoki2021}. UTe$_2$ was first suspected to be a nearly-ferromagnet, in which ferromagnetic fluctuations were thought to lead to triplet superconductivity \cite{Ishizuka2021}. Longitudinal magnetic fluctuations were evidenced by NMR \cite{Tokunaga2019} and muon-spin relaxation measurements \cite{Sundar2019}, but these studies could not unambiguously distinguish ferromagnetic and antiferromagnetic fluctuations. The nature of magnetic order induced under pressure \cite{Braithwaite2019} also constitutes an open question: ferromagnetism was first suggested \cite{Ran2020,Lin2020}, but antiferromagnetism was proposed from more recent studies \cite{Thomas2020,Aoki2020,Aoki2021,Li2021}. In addition, UTe$_2$ crystallizes in an Immm orthorhombic structure, where the U atoms form a two-leg ladder structure with legs along $\mathbf{a}$ and rungs along $\mathbf{c}$ [see Fig. \ref{Fig1}(a-b)] \cite{Hutanu2020,Xu2019,Knafo2021}. Therefore, one could suspect that low dimensionality may impact superconducting pairing too.

Inelastic neutron scattering is wavevector-resolved and allows directly determining whether magnetic fluctuations are ferromagnetic or antiferromagnetic. A first neutron-scattering study of UTe$_2$, by time-of-flight spectroscopy on an assembly of $61$ crystals offering a total mass of 700 mg and mosaicity of $\simeq15^\circ$, led to the identification of antiferromagnetic fluctuations with the incommensurate wavevectors $\mathbf{k}_1=(0,0.57,0)$ and $\mathbf{k}_2=(0,0.43,0)$ [expressed in reciprocal lattice units (rlu) and defined within the first Brillouin zone] \cite{Duan2020}. Here, we present a neutron-scattering study using the triple-axis spectrometer Thales, at the Institut Laue Langevin in Grenoble. A large single crystal of UTe$_2$ with a mass of 241~mg and a mosaicity of $\lesssim2^\circ$ allowed a fine experimental resolution (see Supplementary Materials \cite{SM}). We confirm the presence of magnetic fluctuations with wavevector $\mathbf{k}_1$ and do not see clear signatures of ferromagnetic fluctuations with wavevector $\mathbf{k}=0$. A low-dimensional character related with the ladder structure of the U atoms is emphasized, and the temperature-evolution of the antiferromagnetic fluctuations with wavevector $\mathbf{k}_1$ is carefully investigated.

\begin{figure}[t]
\includegraphics[width=0.9\columnwidth]{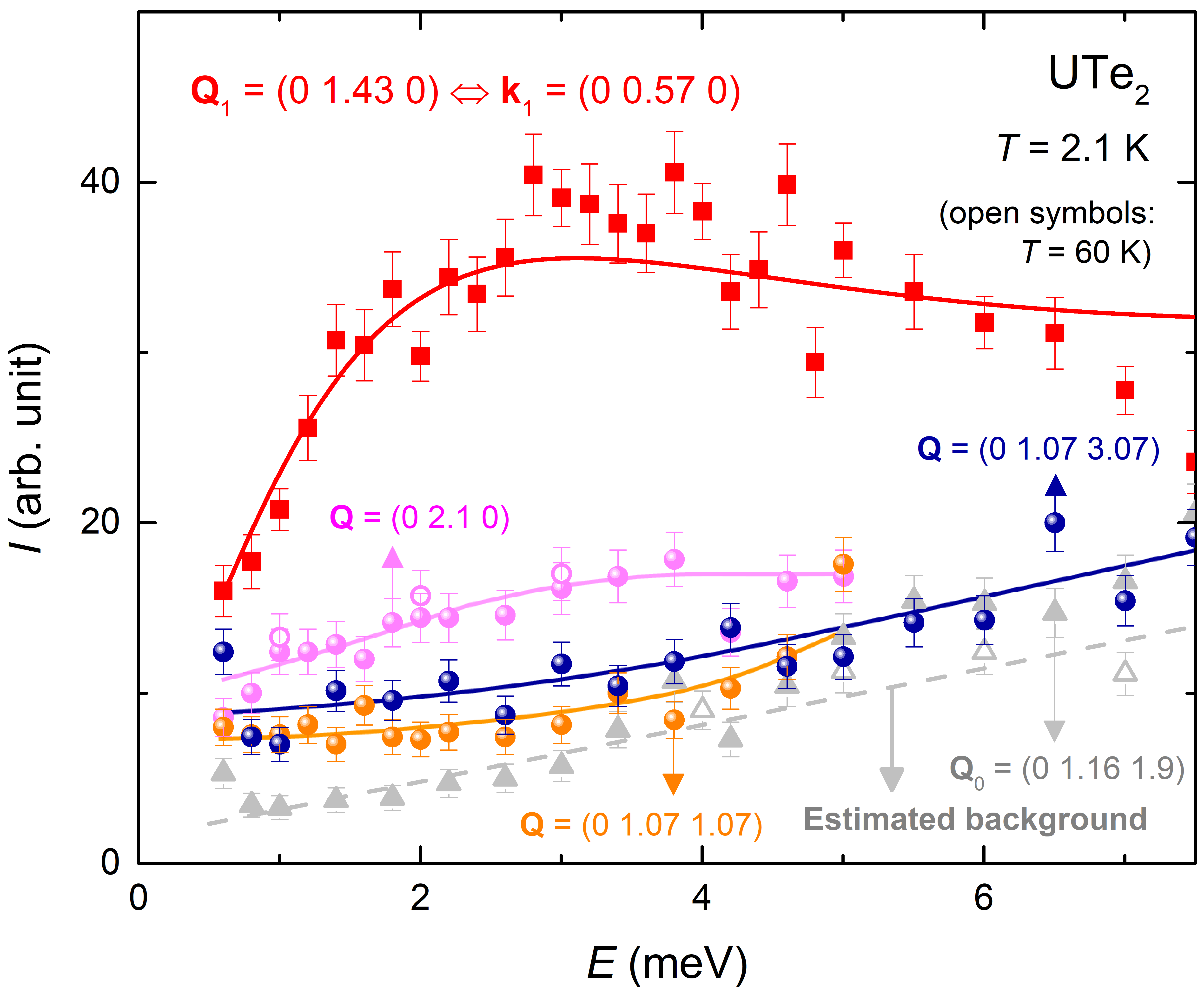}
\caption{Energy scans measured at the temperature $T=2.1$~K and the momentum transfers $\mathbf{Q}_1=(0,1.43,0)$ characteristic of antiferromagnetic fluctuations (red squares), $\mathbf{Q}=(0,2.1,0)$, $\mathbf{Q}=(0,1.07,1.07)$, and $\mathbf{Q}=(0,1.07,3.07)$ characteristic of ferromagnetic fluctuations (circles), and $\mathbf{Q}=(0,1.16,1.9)$ characteristic of the background (gray triangles). A few points are shown at $T=60$~K and the momentum transfers $\mathbf{Q}=(0,2.1,0)$ and $\mathbf{Q}=(0,1.16,1.9)$ (open symbols). Lines correspond to a fit to the data by a quasielastic Lorentzian shape at the momentum transfer $\mathbf{Q}_1$, to a fit to the data by a linear shape at the momentum transfer $\mathbf{Q}=(0,1.16,1.9)$, and to guides to the eyes at the other momentum transfers.}
\label{Fig2}
\end{figure}

Figure \ref{Fig2} presents energy scans measured at different momentum transfers $\mathbf{Q}$ and $T=2.1$~K, for energy transfers $0.6\leq E\leq7.5$~meV. A large signal at the momentum transfer $\mathbf{Q}_1=(0,1.43,0)$ indicates the presence of strong antiferromagnetic fluctuations at the incommensurate wavevector $\mathbf{k}_1=(0,0.57,0)$ [from $\mathbf{Q}_1=\mathbf{\tau}-\mathbf{k}_1$, where $\mathbf{\tau}=(0,2,0)$ is a nuclear Bragg position]. $\mathbf{k}_1$ is close to the Brillouin-zone boundary [see Fig. \ref{Fig1}(c)]. The momentum transfer $\mathbf{Q}=(0,1.16,1.9)$, chosen far from $\mathbf{Q}_1$ (and with the same modulus), is characteristic of a background without magnetic fluctuations. A few points collected at $T=60$~K indicate a nearly-temperature independent background. Spectra at three momentum transfers $\mathbf{Q}=(0,2.1,0)$, $(0,1.07,1.07)$, and $(0,1.07,3.07)$, expected to characterize ferromagnetic fluctuations with wavevector $\mathbf{k}\simeq0$, present intensities near to the background level \footnote{The momentum transfers $\mathbf{Q}=(0,2.1,0)$, $(0,1.07,1.07)$, and $(0,1.07,3.07)$ were chosen near, but not exactly at, the nuclear Bragg positions $\mathbf{\tau}=(0,2,0)$, $(0,1,1)$, and $(0,1,3)$ corresponding to magnetic wavevector $\mathbf{k}=0$ to avoid contamination by the nuclear Bragg peaks.}. At $\mathbf{Q}=(0,2.1,0)$, the intensity is slightly higher than at the two other ferromagnetic positions, but similar intensities measured at $T=2.1$ and 60~K indicate a presumably non-magnetic signal. Within the experimental window investigated here ($T\geq2.1$~K and  $0.6\leq E\leq7.5$~meV), we do not identify, thus, any clear signature of ferromagnetic fluctuations in UTe$_2$.

\begin{figure}[t]
\includegraphics[width=0.9\columnwidth]{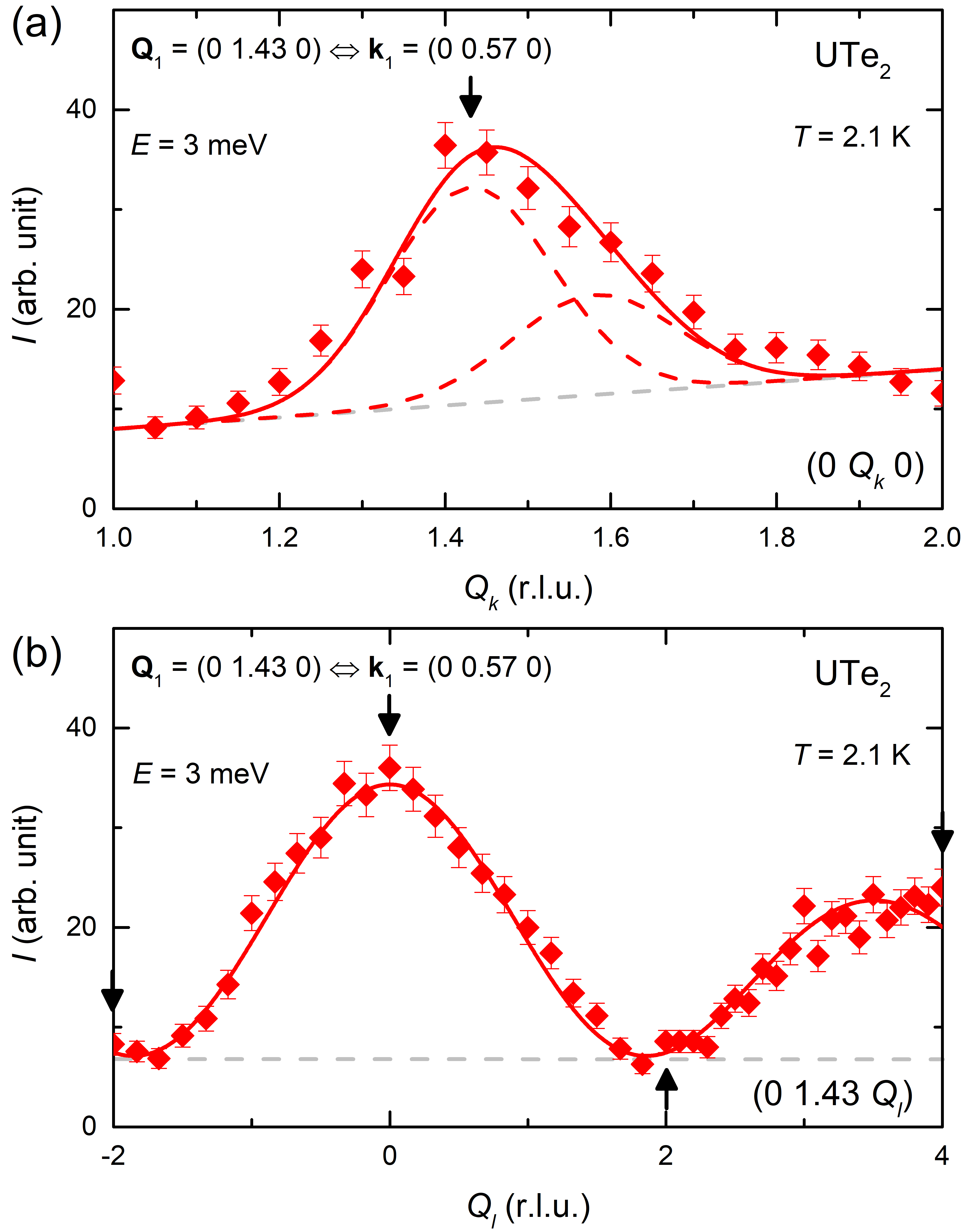}
\caption{(a) $(0,Q_k,0)$ and (b) $(0,1.43,Q_l)$ scans at the energy transfer $E=3$~meV and the temperature $T=2.1$~K. Gray dashed lines correspond to the estimated background. Red lines correspond to fit to the data as described in \cite{SM}.}
\label{Fig3}
\end{figure}

$\mathbf{Q}$-dependences of the antiferromagnetic signal centered at $\mathbf{Q}_1$, measured at the energy transfer $E=3$~meV and at the temperature $T=2.1$~K, are presented in Fig. \ref{Fig3}. The $(0,Q_k,0)$ scan shown in Fig. \ref{Fig3}(a) indicates an asymmetric anomaly of maximal intensity near $\mathbf{Q}_1=(0,1.43,0)$, with a shoulder for $Q_k>1.43$. It can be fitted by two Gaussian contributions of full width at half maximum $\kappa=0.24$~rlu, corresponding to a correlation length $\xi_b=3.75b$, where $b=6.09$~\AA~\cite{Hutanu2020}, along the direction $\mathbf{b}$ (see Supplementary Materials \cite{SM}). The contribution centered at the momentum transfer $\mathbf{Q}_1=(0,1.43,0)$, corresponding to the wavevector $\mathbf{k}_1=(0,0.57,0)$, is twice more intense than that centered at the momentum transfer $\mathbf{Q}_2=(0,1.58,0)$, corresponding to the wavevector $\mathbf{k}_2=(0,0.42,0)$. In the following, focus will be given to the dominant magnetic-fluctuations mode at wavevector $\mathbf{k}_1$. $(0,Q_k,0)$ scans at different energy transfers $E$, shown in the Supplementary Materials \cite{SM}, further indicate that the magnetic fluctuations signal at the wavevector $\mathbf{k}_1=(0,0.57,0)$ is weakly or not dispersive. In Fig. \ref{Fig3}(b), a $(0,1.43,Q_l)$ scan presents a damped sine-wave evolution of the scattered intensity. Best fit to the data is made assuming i) fluctuating magnetic moments $\mathbf{\mu}\parallel\mathbf{a}$ and ii) in-phase  magnetic moments on the two U of the primitive cell [see Fig. \ref{Fig1}(a) and Supplementary Materials \cite{SM}]. The sine-wave modulation indicates that, within first approximation, the correlations along $\mathbf{c}$ can be neglected, i.e., that the inter-ladder magnetic coupling along $\mathbf{c}$ is weak. To our knowledge, magnetic correlations along $\mathbf{a}$ have not been characterized by neutron scattering so far. However, a first component $k_{1,h}=0$ of the propagation vector $\mathbf{k}_1$ is implicitly assumed here and in \cite{Duan2020}. Indeed, strong correlations along the ladder direction $\mathbf{a}$ are expected from the short distance $d_2=4.12$~\AA~between U atoms along $\mathbf{a}$. For these reasons, one can safely expect that the fluctuations with wavevector $\mathbf{k}_1$ imply correlations with in-phase fluctuating moments along $\mathbf{a}$. We can further conclude that the antiferromagnetic fluctuations investigated here are low-dimensional, and that they are characterized by a set of equivalent, or nearly-equivalent, lines of wavevectors $\mathbf{k_L}=(0,0.43,\delta)$ and $(0,0.57,\delta)$, which include $\mathbf{k}_1$ and $\mathbf{k}_2$ (see Supplementary Materials \cite{SM}).

\begin{figure}[t]
\includegraphics[width=0.9\columnwidth]{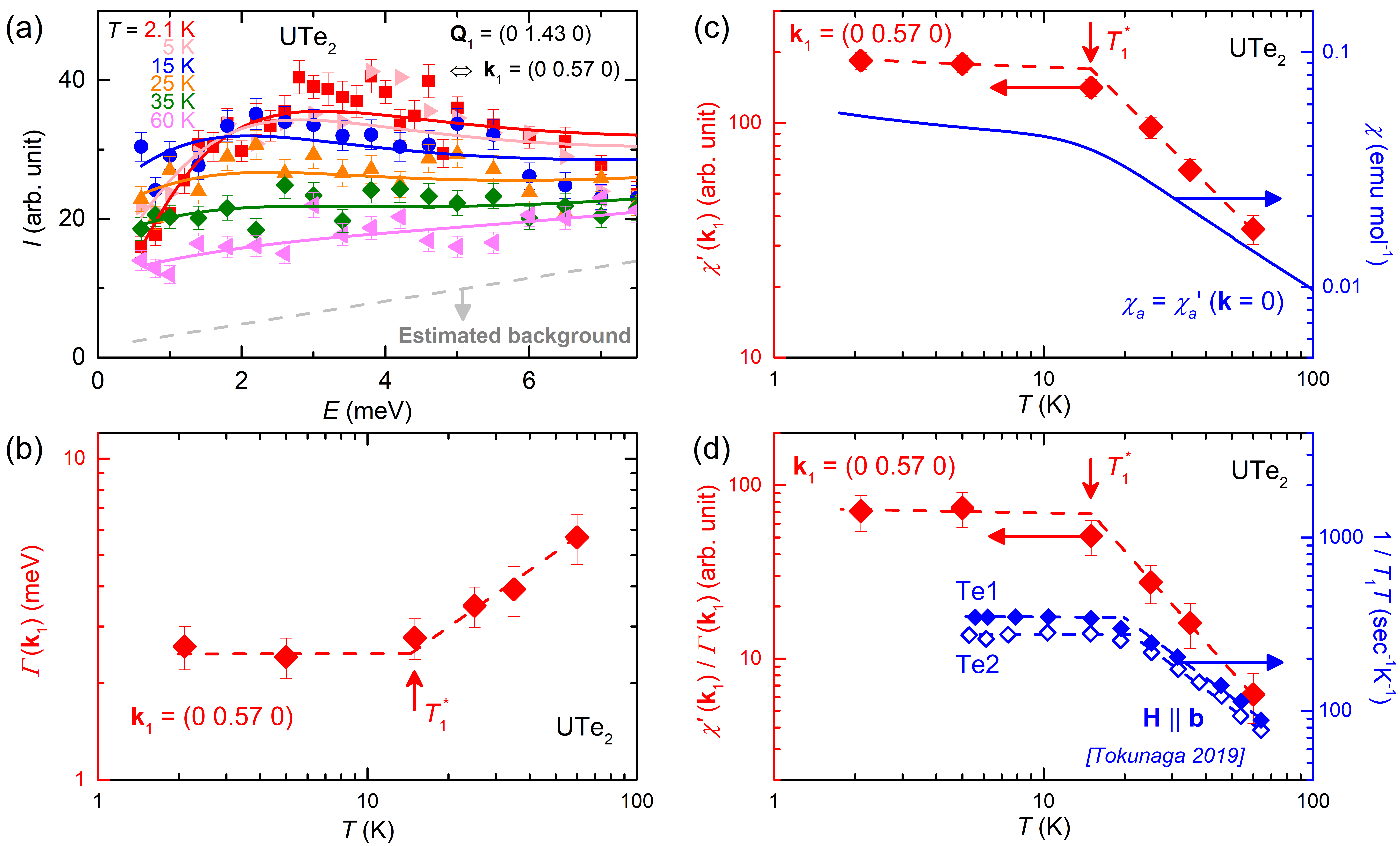}
\caption{(a) Energy scans measured at $\mathbf{Q}_1=(0,1.43,0)$ and temperatures $T$ from $2.1$ to 60~K. Full lines are fits to the data assuming a Lorentzian shape, and the dashed line indicates the estimated background. Temperature dependence, in log-log scales, (b) of the relaxation rate $\Gamma(\mathbf{k}_1)$, (c) of the real part of the static susceptibility $\chi'(\mathbf{k}_1)$ and the bulk magnetic susceptibility $\chi_a=M/H$ measured in a magnetic field $\mu_0\mathbf{H}\parallel\mathbf{a}$ of 0.1~T, and (d) of the ratio $\chi'(\mathbf{k}_1)/\Gamma(\mathbf{k}_1)$ and of the NMR relaxation rate $1/T_1T$ measured on the Te1 and Te2 sites in a magnetic field $\mathbf{H}\parallel\mathbf{b}$ (from \cite{Tokunaga2019}). In panels (c-d), left and right scales have the same number of decades to allow direct comparison between plotted data.}
\label{Fig4}
\end{figure}

Figure \ref{Fig4}(a) shows spectra measured at the momentum transfer $\mathbf{Q}_1=(0,1.43,0)$ for a large set of temperatures $T$ from $2.1$ to 60~K. The increase of $T$ leads to a progressive decrease and broadening of the antiferromagnetic signal. We fitted data shown in Fig. \ref{Fig4}(a) assuming a quasielastic Lorentzian variation of the imaginary part of the dynamical susceptibility $\chi''(\mathbf{Q}_1,E)$ (see Supplementary Materials \cite{SM}). The temperature variations of the relaxation rate $\Gamma(\mathbf{k}_1)$ and of the real part of the static susceptibility $\chi'(\mathbf{k}_1)$ extracted from these fits are plotted within log-log scales in Fig. \ref{Fig4}(b-c). When the temperature is lowered, the strengthening of antiferromagnetic fluctuations is evidenced by the decrease of $\Gamma(\mathbf{k}_1)$ and the increase of $\chi'(\mathbf{k}_1)$, which both saturate at temperatures below $T_1^*\simeq15$~K. $T_1^*$ and the low-temperature value $\Gamma(\mathbf{k}_1)\simeq2.5$~meV~$\simeq 2~k_BT_1^*$ are characteristic of the antiferromagnetic fluctuations with wavevector $\mathbf{k}_1$.

Anomalies are observed at temperatures near $T_1^*$ in various physical properties. The bulk magnetic susceptibility $\chi_a=M/H$, where $M$ is the magnetization and $\mu_0H=0.1$~T a magnetic field applied along the easy magnetic axis $\mathbf{a}$, is compared to $\chi'(\mathbf{k}_1)$ in Fig. \ref{Fig4}(c). Knowing that $\chi_a=\chi_a'(\mathbf{k}=0)$ and $\chi'(\mathbf{k}_1)=\chi_a'(\mathbf{k}_1)$ (since the magnetic fluctuations with wavevector $\mathbf{k}_1$ were attributed to magnetic moments $\mu\parallel\mathbf{a}$), both quantities are expected to converge at temperatures at which intersite magnetic correlations have vanished, i.e., presumably $T\gtrsim100$~K. A broad kink in $\chi_a$ is observed at the temperature $T_{\chi-a}^{kink}\simeq15$~K~$\simeq T_1^*$. Maxima in the electronic heat capacity \cite{Willa2021} and electrical resistivity measured with a current $\mathbf{I}\parallel\mathbf{c}$ \cite{Eo2021} and minima in the thermal expansion measured with lengths $\mathbf{L}\parallel\mathbf{b},\mathbf{c}$ \cite{Thomas2021,Willa2021} and thermoelectric power measured with a current $\mathbf{I}\parallel\mathbf{a}$ \cite{Niu2020} are also observed at a temperature of $\simeq15$~K. These anomalies may result from the development of antiferromagnetic fluctuations with wavevector $\mathbf{k}_1$ and their possible feedbacks on the Fermi surface. Figure \ref{Fig4}(d) presents a comparison of $\chi'(\mathbf{k}_1)/\Gamma(\mathbf{k}_1)$ extracted here and the NMR relaxation rate $1/T_1T$, measured in a magnetic field $\mathbf{H}\parallel\mathbf{b}$ by Tokunaga \textit{et al} (for the two sites Te1 and Te2) \cite{Tokunaga2019}. These two quantities are dominated by fluctuations of magnetic moments $\mathbf{\mu}\parallel\mathbf{a}$ and vary similarly, increasing with decreasing temperatures before saturating at low temperature. $1/T_1T$ saturates below a temperature of $\simeq20$~K near $T_1^*$. Knowing that $1/T_1T$ consists in a sum of $\chi'(\mathbf{k})/\Gamma(\mathbf{k})$ over the reciprocal space (see Supplementary Materials \cite{SM}), the magnetic fluctuations at $\mathbf{k}_1$ (and its equivalent positions $\mathbf{k}_L$) may contribute significantly to $1/T_1T$. The slower $T$-variation of $1/T_1T$, in comparison with that of $\chi'(\mathbf{k}_1)/\Gamma(\mathbf{k}_1)$, may be due to the contribution of fluctuations at wavevectors far from $\mathbf{k}_1$. The question of low-energy ferromagnetic fluctuations (not observed here) and, thus, of their possible contribution to $1/T_1T$, which is sensitive to energies $E\rightarrow0$, remains open.

Similar relationship between anomalies in bulk properties and magnetic fluctuations were observed in other heavy-fermion paramagnets \cite{Aoki2013}. In the prototypical heavy-fermion paramagnet CeRu$_2$Si$_2$, the characteristic temperature $T_1^*\simeq10$~K of longitudinal antiferromagnetic fluctuations \cite{Knafo2009} is comparable to those of anomalies in the magnetic susceptibility \cite{Haen1987}, thermal expansion \cite{Paulsen1990}, and NMR relaxation rate \cite{Ishida1998}. The situation in UTe$_2$ is more complex. In addition to the temperature scale of $\simeq15$~K considered earlier, a second temperature scale of $\simeq35$~K can be identified from maxima in the magnetic susceptibility $\chi_b$ measured with $\mathbf{H}\parallel\mathbf{b}$ (hard magnetic axis) \cite{Ikeda2006}, in the Hall coefficient \cite{Niu2020} and electrical resistivity \cite{Knafo2019} measured with a current $\mathbf{I}\parallel\mathbf{a}$. Further experiments are needed to determine the relationship between the 15-K and 35-K anomalies, and to test if a second and higher-energy magnetic-fluctuations mode could be present.

The possible coexistence of ferromagnetic and antiferromagnetic couplings in the ladder structure of UTe$_2$ was theoretically discussed in \cite{Xu2019}, and a first report of antiferromagnetic fluctuations was made in \cite{Duan2020}. Thanks to a careful investigation of this antiferromagnetic-fluctuations mode, we emphasize here its quasielastic and low-dimensional characteristics. It can be visualized as fluctuations of magnetic moments $\mathbf{\mu}\parallel\mathbf{a}$, with the following phenomenological scheme: i) an inter-ladder antiferromagnetic coupling $J^b_{inter}(AF)$ ending in sine-wave modulation with wavevector $\mathbf{k}_1$ of the moments, ii) intra-ladder ferromagnetic couplings $J^c_1(FM)$ along $\mathbf{c}$ and $J^a_2(FM)$ along $\mathbf{a}$, and iii) a weak inter-ladder magnetic coupling $J^c_{inter}$ along $\mathbf{c}$ [see Fig. \ref{Fig1}(a)] \footnote{The phenomenological inter-ladder magnetic couplings $J^b_{inter}$ and $J^c_{inter}$ introduced here may result from two successive 'diagonal' inter-ladder exchanges [along third shortest U-U distance $d_3$, see Fig. \ref{Fig1}(a-b)].}. $\mathbf{Q}$-modulations of a magnetic-fluctuations signal similar to those reported here for UTe$_2$ were observed by inelastic neutron scattering in other low-dimensional magnets whose primitive cell contains several magnetic ions, as the magnetic ladder Sr$_{14}$Cu$_{24}$O$_{41}$ \cite{Regnault1999,Boullier2005}, the layered paramagnet Sr$_3$Ru$_2$O$_7$ \cite{Capogna2003}, antiferromagnet YBa$_2$Cu$_3$O$_{6.2}$ \cite{Sato1988}, and superconductors YBa$_2$Cu$_3$O$_{6.85}$ \cite{Pailhes2004} and CaKFe$_4$As$_4$ \cite{Xie2018}. It has been argued that a more robust superconducting pairing is expected for two-dimensional, rather than three-dimensional, itinerant magnets \cite{Monthoux2001}. A low-dimensional character of the magnetic fluctuations may, thus, be of importance for the development of superconductivity in UTe$_2$. The question of the role of ferromagnetic fluctuations for the development of a superconducting phase, often suspected to be triplet, needs to be clarified. Ferromagnetic fluctuations were not observed here, within the investigated experimental window $0.6\leq E\leq7.5$~meV, at $T=2.1$~K, but Duan \textit{et al} reported a small signal at the ferromagnetic momentum transfer $\mathbf{Q}=(0,1,1)$, at $E\simeq0.4$~meV and $T=300$~mK \cite{Duan2020}. An unambiguous experimental evidence for ferromagnetic fluctuations is now needed. We note that in the paramagnet Sr$_2$RuO$_4$, a former candidate for triplet superconductivity, intense incommensurate antiferromagnetic fluctuations were found to coexist with weak, and broad in $\mathbf{Q}$-space, quasi-ferromagnetic fluctuations \cite{Pustogow2019,Steffens2019}. Conversely, triplet superconductivity was proposed to occur in UPt$_3$, where only antiferromagnetic fluctuations were observed \cite{Aeppli1988,Tou1998}. Our finding that the antiferromagnetic fluctuations imply ferromagnetically-coupled U atoms on the ladders may be of importance. Indeed, it is compatible with the proposition of pseudo-triplet superconductivity induced by a ferromagnetic interaction between U atoms of the ladder rungs, without necessarily implying ferromagnetic fluctuations \cite{Xu2019,Shishidou2021,Anderson1985}. In the future, challenges will be to determine precisely the respective roles of antiferromagnetic and ferromagnetic fluctuations and low dimensionality for the stabilization of different superconducting phases in UTe$_2$. The topology and possible low-dimensionality of some Fermi-surface sheets, which set in at low temperatures, also play a role for superconductivity \cite{Xu2019,Miao2020,Shishidou2021} and would merit further consideration too.

\onecolumngrid

\twocolumngrid

\vspace{2cm}

\section*{Acknowledgements}

We acknowledge useful discussions with Y. Tokunaga, K. Ishida, and C. Simon. \\

This work was supported by the ANR FRESCO.

\onecolumngrid

\renewcommand{\theequation}{S\arabic{equation}}
\renewcommand{\thefigure}{S\arabic{figure}}
\renewcommand{\bibnumfmt}[1]{[S#1]}
\renewcommand{\citenumfont}[1]{S#1}
 \setcounter{figure}{0}

\vspace{15cm}
\begin{center}
\large {\textbf {Supplementary Materials:\\ Low-dimensional antiferromagnetic fluctuations in the heavy-fermion paramagnetic ladder UTe$_2$}}
\end{center}
\vspace{1cm}

In these Supplementary Materials, we present complementary graphs and details concerning the analysis of our data obtained by inelastic neutron scattering on a single crystal of UTe$_2$.

\section{Sample growth and characterization}

\begin{figure}[b]
\includegraphics[width=.7\columnwidth]{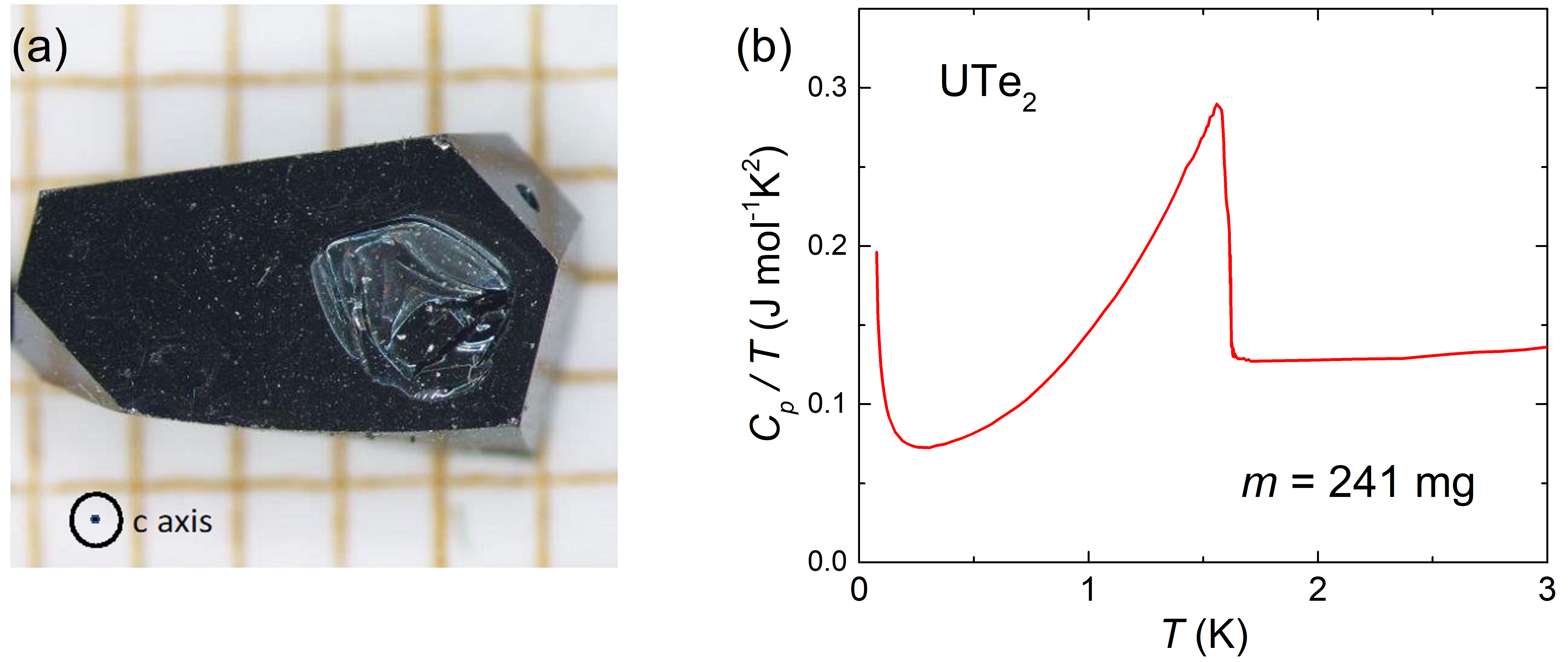}
\caption{(a) Photo and (b) heat capacity $C_p$, plotted as $C_p/T$ versus $T$, of the single crystal used in the present neutron scattering study.}
\label{FigS1}
\end{figure}

The 241-mg single-crystalline sample of UTe$_2$ used in this neutron-scattering study was grown using chemical vapor transport method. Figure \ref{FigS1}(a) presents a photo of the crystal. Its heat-capacity $C_p$, measured at the CEA-Grenoble, is presented as $C_p/T$ versus $T$ graph in Fig. \ref{FigS1}(b). The onset of superconductivity is visible at the temperature $T_{sc}\simeq1.6$~K. The transition is sharp, with a total width $\Delta T\simeq0.05$~K. This indicates a high quality and small degree of distortion of the crystal. A large increase of $C_p/T$ is also observed at temperatures $T<250$~mK.

\begin{figure}[t]
\includegraphics[width=.5\columnwidth]{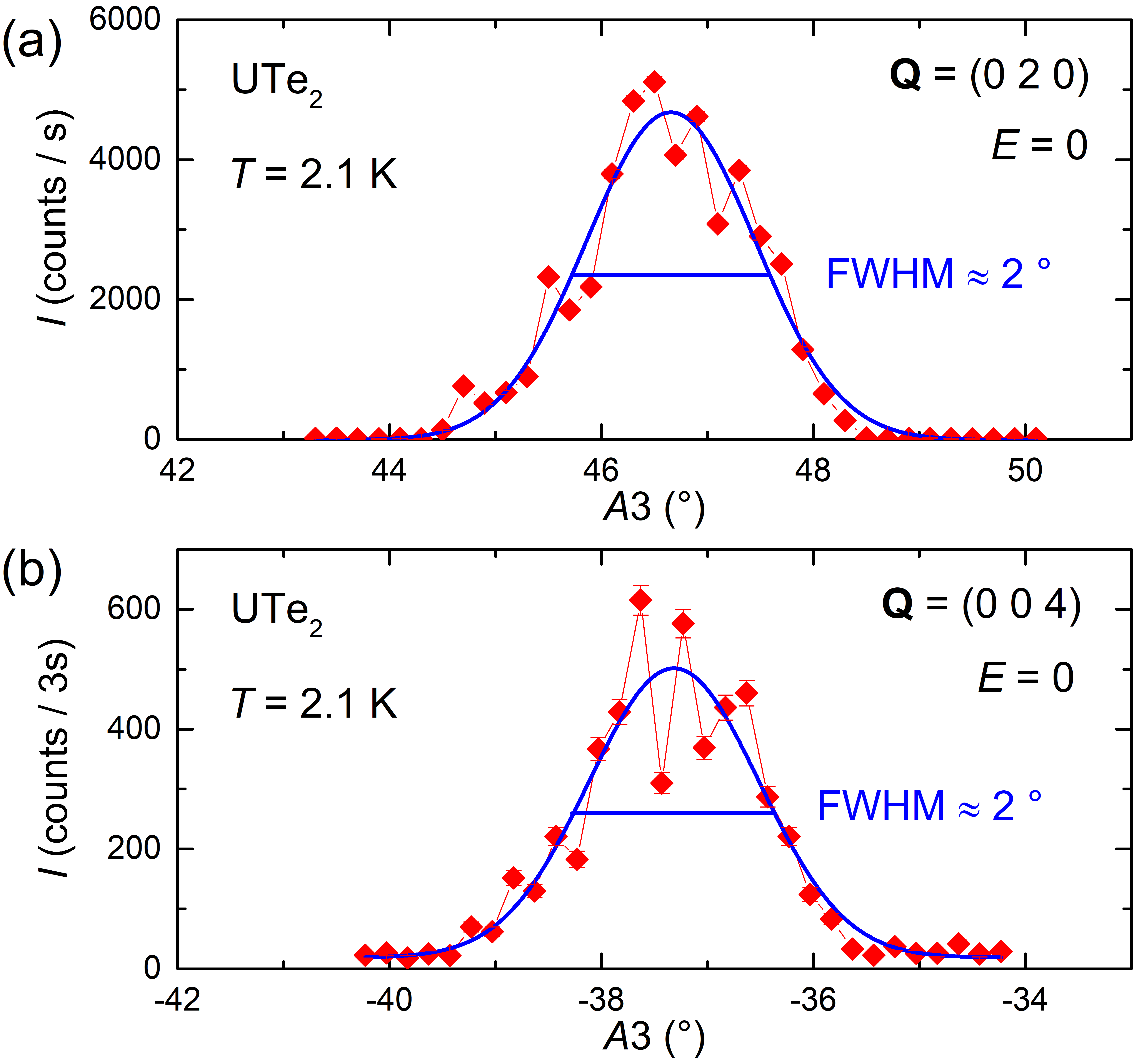}
\caption{$A3$-scans around the Bragg positions $\mathbf{Q}=(0,2,0)$ and $\mathbf{Q}=(0,0,4)$ measured by neutron diffraction at the energy $E=0$ and the temperature $T=2$~K.}
\label{FigS2}
\end{figure}

$A3$-scans, where the angle $A3$ characterizes the rotation of the sample relatively to the vertical axis, around the Bragg positions $\mathbf{Q}=(0,2,0)$ and $\mathbf{Q}=(0,0,4)$ and measured by elastic neutron diffraction at the temperature $T=2$~K, are presented in Fig. \ref{FigS2}. These scans show that the sample has a mosaicity of $\lesssim2~^\circ$.

\section{$\mathbf{Q}$-scans at different energies}

\begin{figure}[t]
\includegraphics[width=.5\columnwidth]{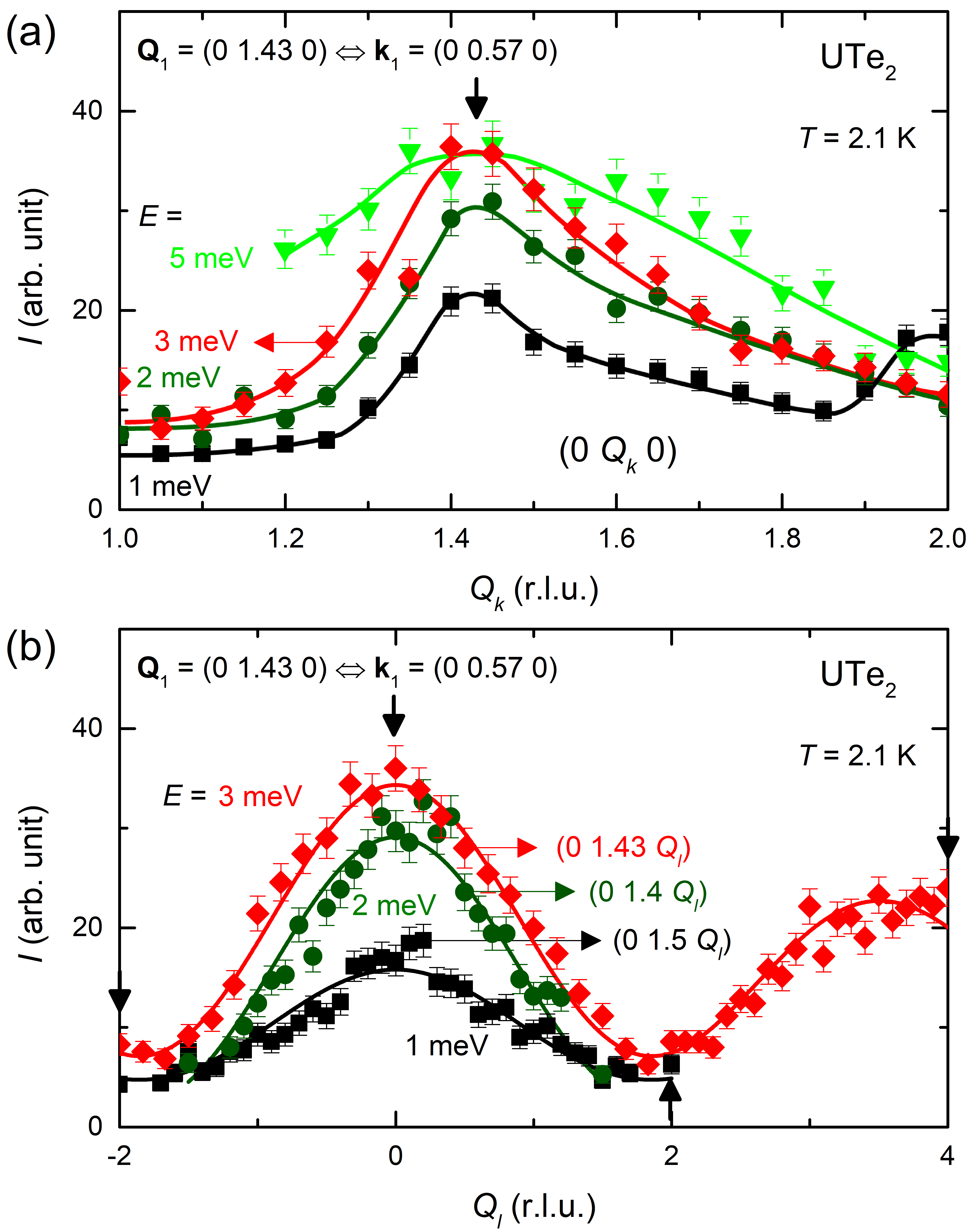}
\caption{(a) $(0,Q_k,0)$ scans at the energy transfers $E=1,3,5$~meV and the temperature $T=2.1$~K. Full lines are guides to the eyes. (b) $(0,Q_k,Q_l)$ scans, where $Q_k$ is fixed to a value $1.4\leq Q_k\leq1.5$, at the energy transfers $E=1,3,5$~meV and the temperature $T=2.1$~K. Full lines correspond to fits to the data.}
\label{FigS3}
\end{figure}

Complementarily to Fig. 3 of the paper, Fig. \ref{FigS3} presents $\mathbf{Q}$-scans around the antiferromagnetic signal centered at $\mathbf{Q}_1=(0,1.43,0)$, measured at energy transfers $E=1,3,~\rm{and}~5$~meV and at the temperature $T=2.1$~K. The $(0,Q_k,0)$ scans shown in Fig. \ref{FigS3}(a) present an asymmetric anomaly of maximal intensity near $\mathbf{Q}_1=(0,1.43,0)$, with a shoulder for $Q_k>1.43$. The $(0,Q_k,Q_l)$ scans at constant $Q_k$ shown in Fig. \ref{FigS3}(b) show a damped sine-wave variation of intensity. These $\mathbf{Q}$ scans performed at different energies indicate that the excitations peaked at $\mathbf{Q}_1$ form a rod as a function of energy, and is therefore weakly or not dispersive in the energy range investigated here.

$Q_k$ scans were fitted to Gaussian contributions expressed as:
\begin{eqnarray}
I(Q_k)=I_0\exp[-4\ln2(Q_k/\kappa)^2],
\label{Gauss_Qk}
\end{eqnarray}
where $\kappa$ corresponds to the full-width at half maximum (FWHM) of the contribution. A Fourier transformation of Eq. (\ref{Gauss_Qk}) leads to the real-space distribution of intensity:
\begin{eqnarray}
I'(y)=I_0'\exp[-4\ln2(y/\xi_b)^2],
\label{Gauss_y}
\end{eqnarray}
where $\xi_b=4b\ln2/(\pi\kappa)$ is defined as the correlation length in the direction $\mathbf{b}$ of the magnetic fluctuations considered here.

\begin{figure}[t]
\includegraphics[width=.95\columnwidth]{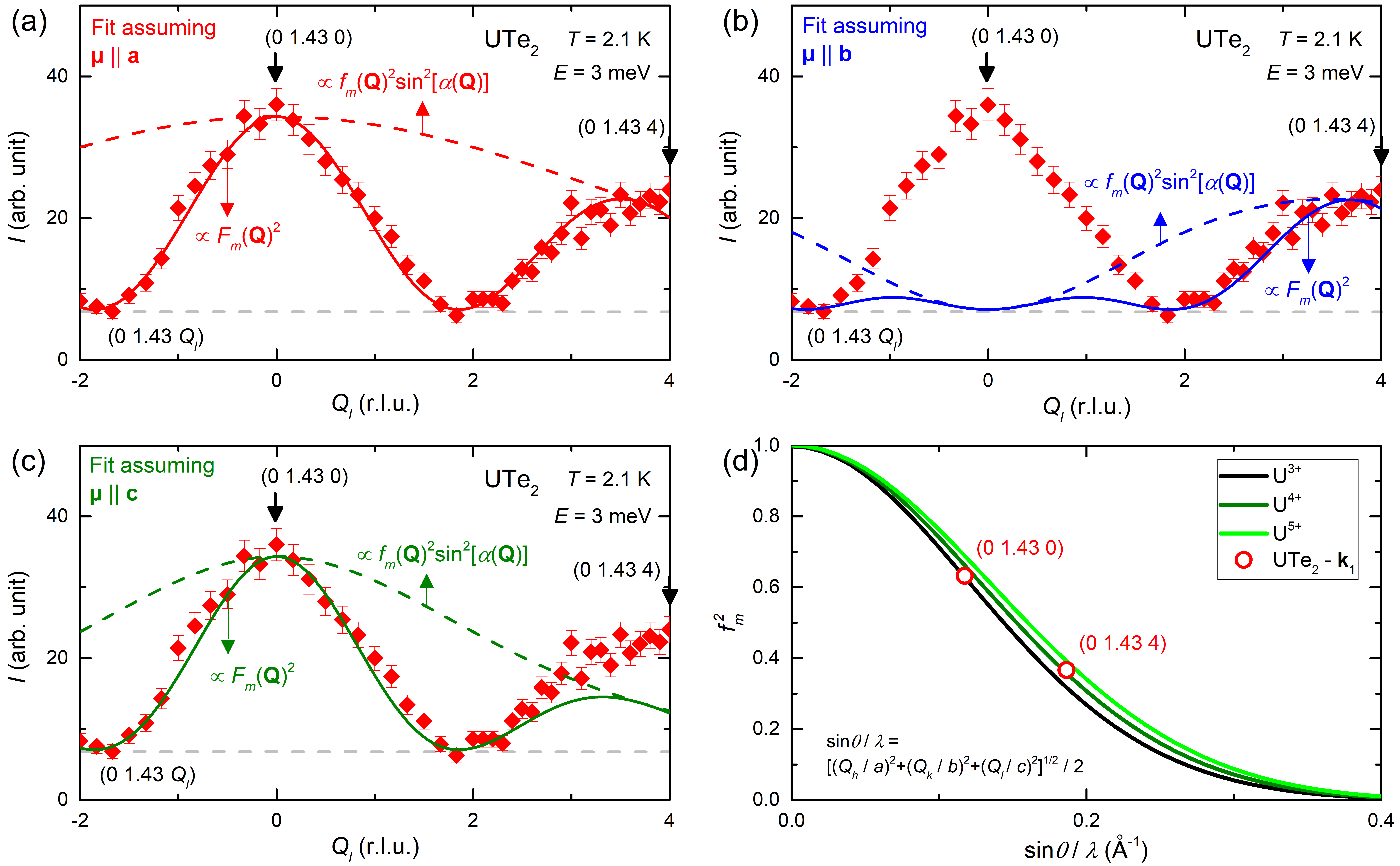}
\caption{$(0,1.43,Q_l)$ scan at the energy transfer $E=3$~meV and the temperature $T=2.1$~K and its fits assuming magnetic moments (a) $\mathbf{\mu}\parallel\mathbf{a}$, (b) $\mathbf{\mu}\parallel\mathbf{b}$, and (c) $\mathbf{\mu}\parallel\mathbf{c}$. Dashed gray lines correspond to the estimated background, dashed colored lines to the envelope $f_m(\mathbf{Q})^2\sin^2[\alpha(\mathbf{Q})]$, and full colored lines correspond to fit by $F_m(\mathbf{Q})^2$ to the data. (d) Variation of magnetic form factor $f_m$ with $\sin\theta/\lambda$, where $\theta$ is the scattering angle and $\lambda$ the neutron wavelength, expected for U$^{3+}$, U$^{4+}$, and U$^{5+}$ [\onlinecite{Brown2002}], and comparison with scaled intensities measured on UTe$_2$ at the momentum transfers $(0,1.43,0)$ and $(0,1.43,4)$, for $E=3$~meV and $T=2.1$~K.}
\label{FigS4}
\end{figure}

\begin{figure}[t]
\includegraphics[width=1\columnwidth]{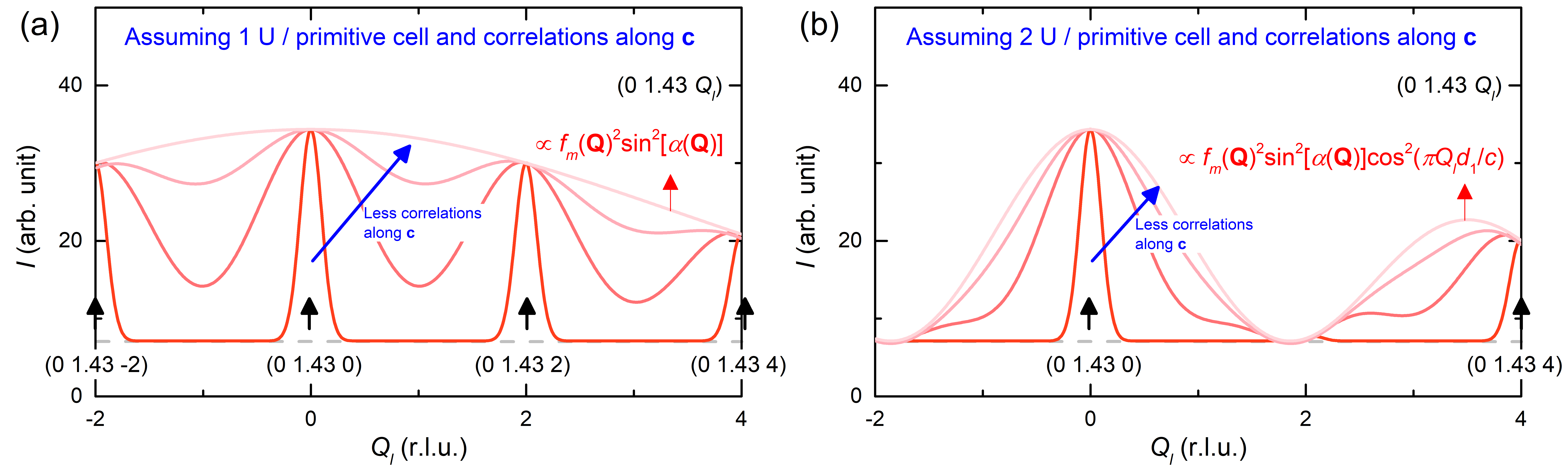}
\caption{$Q_l$ scans expected for different strengths (from strong to weak) of correlations along the direction $\mathbf{c}$, assuming (a) one U atom per primitive cell (b) in phase magnetic moments on the two U atoms of the primitive cell of UTe$_2$.}
\label{FigS5}
\end{figure}

\section{Fit to the $(0,1.43,Q_l)$ scan by a damped sine-wave function}
\label{section_fit}

Figure \ref{FigS4} presents details concerning the fit to the damped sine-wave variation of intensity from the $(0,1.43,Q_l)$ scan made at $T=2.1$~K and shown in Fig. 3(b) of the paper. Data were fitted by a function proportional to the square $F_m(\mathbf{Q})^2$ of the magnetic structure factor:
\begin{eqnarray}
F_m(\mathbf{Q})^2=f_m(\mathbf{Q})^2\sin^2[\alpha(\mathbf{Q})]\cos^2(\pi Q_l d_1/c),
\label{fit_QL}
\end{eqnarray}
where $f_m(\mathbf{Q})$ is the magnetic form factor of individual U atoms, $\alpha(\mathbf{Q})$ is the angle between $\mathbf{Q}$ and the magnetic moment $\mu$, and the cosine pre-factor is controlled by the magnetic pattern in the primitive cell, with $d_1=3.72$~\AA~[see Fig. 1(a) of the paper]. The envelope $f_m(\mathbf{Q})^2\rm{sin}^2[\alpha(\mathbf{Q})]$ and the full fitted function $F_m(\mathbf{Q})^2$ are plotted assuming the three hypotheses $\mathbf{\mu}\parallel\mathbf{a}$, $\mathbf{\mu}\parallel\mathbf{b}$, and $\mathbf{\mu}\parallel\mathbf{c}$ in Fig. \ref{FigS4}(a), (b) and (c), respectively. While a good fit to the data is obtained assuming $\mathbf{\mu}\parallel\mathbf{a}$, the hypotheses $\mathbf{\mu}\parallel\mathbf{b}$ and $\mathbf{\mu}\parallel\mathbf{c}$ do not permit fitting the data in the whole measured $Q_l$ range. In this fit, the magnetic form factor $f_m(\mathbf{Q})$ of valence-3 U$^{3+}$ ions was used [\onlinecite{Brown2002}], following the conclusions from a photoelectron-spectroscopy study [\onlinecite{Fujimori2021}]. We note that a valence 4 corresponding to U$^{4+}$ ions was also proposed from an x-ray-absorption-spectroscopy experiment and the question of valence in UTe$_2$ remains open [\onlinecite{Thomas2020}]. However, the fit done here is almost unaffected by the choice of valence, since the magnetic form factors of U$^{3+}$, U$^{4+}$, but also U$^{5+}$ ions are very similar [\onlinecite{Brown2002}]. The pre-factor $\rm{cos}(\pi Q_l d_1/c)$, which allows a good fit to the data, is expected for in-phase, i.e., ferromagnetically-coupled, magnetic moments on the two U of the primitive cell. We note that the first zero of the pre-factor occurs for $Q_l=c/(2d_1)$, which is accidentally close to $\simeq2$. Alternative hypotheses can lead to different pre-factors, as $\rm{sin}(\pi Q_l d_1/c)$ expected for antiparallel magnetic moments, but do not reproduce the experimental data.

For pedagogical purpose, Fig. \ref{FigS5} shows $Q_l$ scans expected for different strengths of the correlations along the direction $\mathbf{c}$, assuming one U atom per primitive cell [Fig. \ref{FigS5}(a)] or assuming in phase magnetic moments on the two U atoms of the primitive cell of UTe$_2$ [Fig. \ref{FigS5}(b)]. Well-defined peaks in the case of strong correlations are replaced by broader peaks when the correlations are weakened, ending asymptotically in the envelope given by Eq. (\ref{fit_QL}) when the correlations vanish.

\section{Magnetic fluctuations with wavevectors $\mathbf{k}_1$ and $\mathbf{k}_2$}

\begin{figure}[t]
\includegraphics[width=0.9\columnwidth]{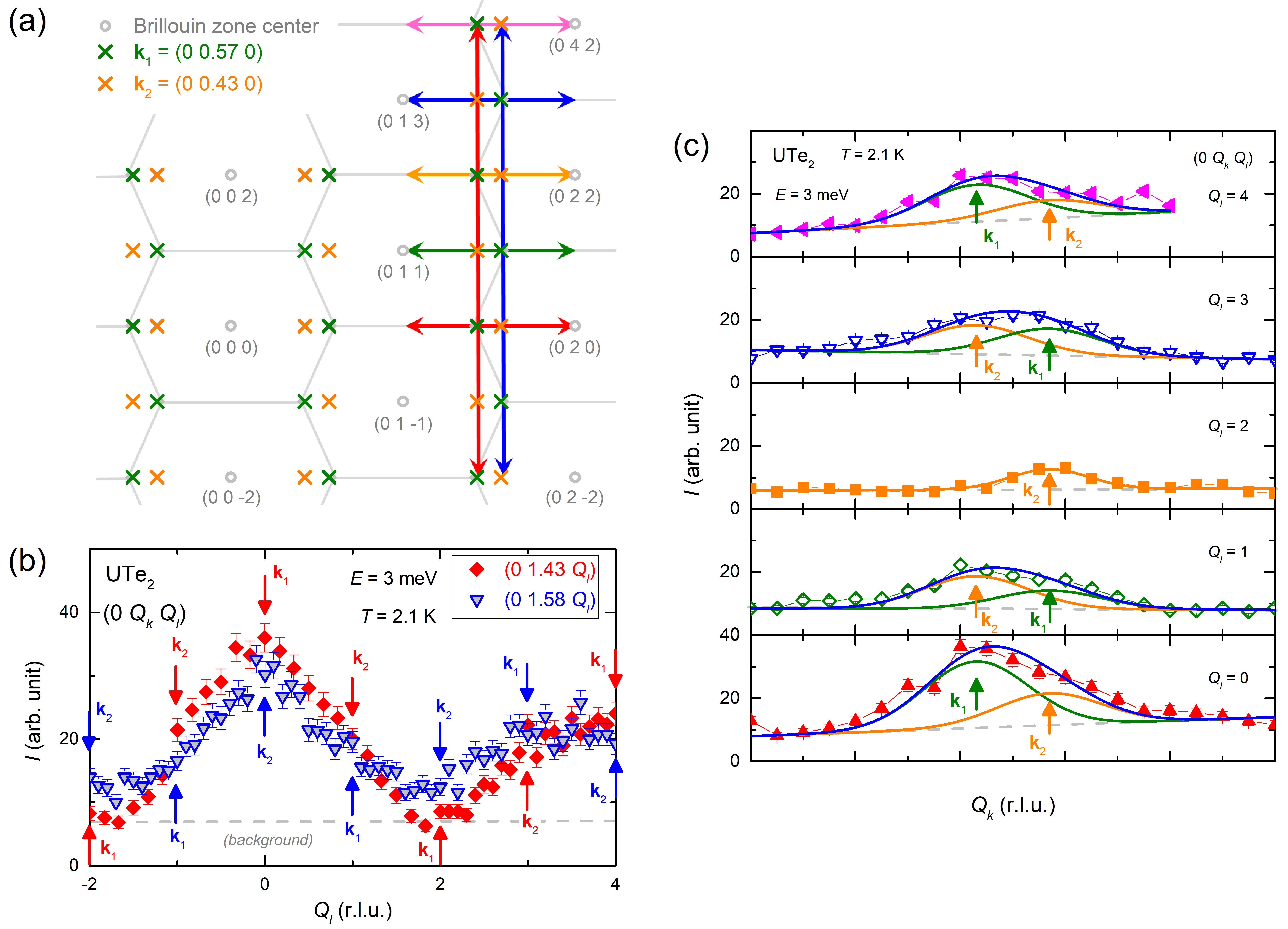}
\caption{(a) Scattering plane of the momentum transfers $Q=(0,Q_k,Q_l)$ accessed in the neutron experiment. Points at the Brillouin zone centers corresponding to wavevector $\mathbf{k}=0$ are indicated by gray open circles. Points corresponding to the wavevectors $\mathbf{k}_1=(0,0.57,0)$ and $\mathbf{k}_2=(0,0.42,0)$ are indicated by green and orange crosses, respectively. $\mathbf{Q}$ scans shown in panels (b-c) are indicated by arrows. (b) $(0,1.43,Q_l)$ and  $(0,1.58,Q_l)$ scans and (c) $(0,Q_k,Q_l)$ scans, with $Q_l=0,1,2,3,4$, performed at the energy $E=3$~meV and the temperature $T=2.1$~K. In panel (c), the lines correspond to fit to the data by one or two Gaussian contributions.}
\label{FigS6}
\end{figure}

Figure \ref{FigS6} presents series of $\mathbf{Q}$-scans performed at the energy $E=3$~meV and the temperature $T=2.1$~K. These scans are indicated by arrows in the scattering plane shown in Fig. \ref{FigS5}(a). The $(0,1.43,Q_l)$ and $(0,1.58,Q_l)$ scans shown in Fig. \ref{FigS6}(b) both present a damped sine-wave evolution of the intensity. Each of these scans presents an alternation of $\mathbf{k}_1$ and $\simeq\mathbf{k}_2$ positions. Within first approximation, their fit to Eq. \ref{fit_QL} (see Sect. \ref{section_fit}) indicates that the antiferromagnetic fluctuations are characterized by a set of wavevectors $\mathbf{k_L}=(0,0.43,\delta)$ and $(0,0.57,\delta)$, which include $\mathbf{k}_1$ and $\mathbf{k}_2$. These two lines of wavevectors are equivalent for symmetry reasons. However, small deviations are indicated by different scattered intensities at positions corresponding to $\mathbf{k}_1$ and $\mathbf{k}_2$. These deviations are clear in Fig. \ref{FigS6}(c), which presents $(0,Q_k,Q_l)$ scans at constant $Q_l=0,1,2,3,4$. Fits made using Gaussian contributions of width $\kappa=0.19-0.24$~rlu and centered at positions corresponding to the wavevectors $\mathbf{k}_1=(0,0.57,0)$ and $\mathbf{k}_2=(0,0.43,0)$ are indicated by full lines in the graphs. These $(0,Q_k,Q_l)$ scans (for constant $Q_l$ values) show a non-trivial evolution of the two peak intensities:
\begin{itemize}
  \item the intensity at wavevector $\mathbf{k}_2$ probed at the momentum transfer $\mathbf{Q}=(0,1.57,0)$ is smaller than that at the wavevector $\mathbf{k}_1$ probed at the momentum transfer $\mathbf{Q}=(0,1.43,0)$ (they should be nearly equal),
  \item a small intensity remains at wavevector $\mathbf{k}_2$ probed at the momentum transfer $\mathbf{Q}=(0,1.43,2)$.
\end{itemize}
The observed intensities are not compatible with simple phonon contaminations. These deviations may be due to subtle magneto-elastic effects. For instance, the acoustic phonon intensity is expected to be one or two orders of magnitude higher for $Q_k=2$ than for $Q_k=1$, which may lead to observable coupling with magnetism. A more complete set of experiments, i.e., energy scans at $\mathbf{k}_2$ positions and momentum-transfer scans at different energies, possibly at different temperatures, may be needed to study the origin of these effects. Great care to the $\mathbf{Q}$-dependence of the experimental background and of the experimental resolution may be mandatory for such works, for which a larger crystal than the one used here may be needed.

We note that, in their work made on an assembly of 61 single crystals using a time-of-flight spectrometer, Duan et \textit{al.} observed a vanishing of magnetic fluctuations signal in the line $(0,Q_k,2)$, in agreement with the cosine term in Eq. (\ref{fit_QL}) used to fit our data [\onlinecite{Duan2020}].

\section{Fit to the energy scans by a quasielastic Lorentzian dynamical susceptibility}

The complex magnetic susceptibility, expressed as:
\begin{eqnarray}
\chi(\mathbf{k},E)=\chi'(\mathbf{k},E)+i\chi''(\mathbf{k},E),
\label{chi}
\end{eqnarray}
where $\chi'(\mathbf{k},E)$ is its real part and $\chi''(\mathbf{k},E)$ its imaginary part, is generally used to described magnetic fluctuations in the reciprocal space.

Knowing that the neutron intensity $I(\mathbf{Q},E)$ is related to the imaginary part of the dynamical susceptibility $\chi''(\mathbf{Q},E)$ by:
\begin{eqnarray}
I(\mathbf{Q},E)-I_{BG}(E)\propto\frac{F_m(\mathbf{Q})^2}{1-\exp(-E/k_BT)}\chi''(\mathbf{Q},E),
\label{I_neutron}
\end{eqnarray}
where $I_{BG}(E)$ is the background, data shown in Fig. 4(a) of the paper were fitted assuming a quasielastic Lorentzian variation of $\chi''(\mathbf{Q}_1,E)$:
\begin{eqnarray}
\chi''(\mathbf{Q}_1,E)=\chi'(\mathbf{k}_1)\frac{E/\Gamma(\mathbf{k}_1)}{1+(E/\Gamma(\mathbf{k}_1))^2},
\label{fit_QE}
\end{eqnarray}
where $\Gamma(\mathbf{k}_1)$ and $\chi'(\mathbf{k}_1)$ are the relaxation rate and the real part of the static susceptibility, respectively, at the wavevector $\mathbf{k}_1$ measured at the momentum transfer $\mathbf{Q}_1$.
We note that a quasielastic Lorentzian variation implies:
\begin{eqnarray}
\chi(\mathbf{k},E)=\frac{\chi'(\mathbf{k})}{1-iE/\Gamma(\mathbf{k})},
\label{chi_QE}
\end{eqnarray}
and that the real part of static susceptibility is given by:
\begin{eqnarray}
\chi'(\mathbf{k})=\chi'(\mathbf{k},E=0)=\frac{1}{\pi}\int_{-\infty}^{\infty}\frac{\chi''(\mathbf{k})}{E}dE.
\label{chi_QE}
\end{eqnarray}

For a $\mathbf{k}$-dependent quasielastic-Lorentzian magnetic susceptibility, the NMR spin-lattice relaxation rate $1/T_1T$ is related to $\chi'(\mathbf{k})$ and $\Gamma(\mathbf{k})$ by:
\begin{eqnarray}
\frac{1}{T_1T}\propto\sum_\mathbf{k}|A(\mathbf{k})|^2\frac{\chi'(\mathbf{k})}{\Gamma(\mathbf{k})},
\label{NMR_1sT1T}
\end{eqnarray}
where $A(\mathbf{k})$ is a $\mathbf{k}$-dependent hyperfine coupling. A strong $\mathbf{k}$ dependence of the weighting factor $|A(\mathbf{k})|^2$ can enhance the contribution at given wavevectors.


\begin{thebibliography}{58}%
\makeatletter
\providecommand \@ifxundefined [1]{%
 \@ifx{#1\undefined}
}%
\providecommand \@ifnum [1]{%
 \ifnum #1\expandafter \@firstoftwo
 \else \expandafter \@secondoftwo
 \fi
}%
\providecommand \@ifx [1]{%
 \ifx #1\expandafter \@firstoftwo
 \else \expandafter \@secondoftwo
 \fi
}%
\providecommand \natexlab [1]{#1}%
\providecommand \enquote  [1]{``#1''}%
\providecommand \bibnamefont  [1]{#1}%
\providecommand \bibfnamefont [1]{#1}%
\providecommand \citenamefont [1]{#1}%
\providecommand \href@noop [0]{\@secondoftwo}%
\providecommand \href [0]{\begingroup \@sanitize@url \@href}%
\providecommand \@href[1]{\@@startlink{#1}\@@href}%
\providecommand \@@href[1]{\endgroup#1\@@endlink}%
\providecommand \@sanitize@url [0]{\catcode `\\12\catcode `\$12\catcode
  `\&12\catcode `\#12\catcode `\^12\catcode `\_12\catcode `\%12\relax}%
\providecommand \@@startlink[1]{}%
\providecommand \@@endlink[0]{}%
\providecommand \url  [0]{\begingroup\@sanitize@url \@url }%
\providecommand \@url [1]{\endgroup\@href {#1}{\urlprefix }}%
\providecommand \urlprefix  [0]{URL }%
\providecommand \Eprint [0]{\href }%
\providecommand \doibase [0]{http://dx.doi.org/}%
\providecommand \selectlanguage [0]{\@gobble}%
\providecommand \bibinfo  [0]{\@secondoftwo}%
\providecommand \bibfield  [0]{\@secondoftwo}%
\providecommand \translation [1]{[#1]}%
\providecommand \BibitemOpen [0]{}%
\providecommand \bibitemStop [0]{}%
\providecommand \bibitemNoStop [0]{.\EOS\space}%
\providecommand \EOS [0]{\spacefactor3000\relax}%
\providecommand \BibitemShut  [1]{\csname bibitem#1\endcsname}%
\let\auto@bib@innerbib\@empty
\bibitem [{\citenamefont {Ran}\ \emph {et~al.}(2019{\natexlab{a}})\citenamefont
  {Ran}, \citenamefont {Eckberg}, \citenamefont {Ding}, \citenamefont
  {Furukawa}, \citenamefont {Metz}, \citenamefont {Saha}, \citenamefont {Liu},
  \citenamefont {Zic}, \citenamefont {Kim}, \citenamefont {Paglione},\ and\
  \citenamefont {Butch}}]{Ran2019}%
  \BibitemOpen
  \bibfield  {author} {\bibinfo {author} {\bibfnamefont {S.}~\bibnamefont
  {Ran}}, \bibinfo {author} {\bibfnamefont {C.}~\bibnamefont {Eckberg}},
  \bibinfo {author} {\bibfnamefont {Q.-P.}\ \bibnamefont {Ding}}, \bibinfo
  {author} {\bibfnamefont {Y.}~\bibnamefont {Furukawa}}, \bibinfo {author}
  {\bibfnamefont {T.}~\bibnamefont {Metz}}, \bibinfo {author} {\bibfnamefont
  {S.~R.}\ \bibnamefont {Saha}}, \bibinfo {author} {\bibfnamefont {I.-L.}\
  \bibnamefont {Liu}}, \bibinfo {author} {\bibfnamefont {M.}~\bibnamefont
  {Zic}}, \bibinfo {author} {\bibfnamefont {H.}~\bibnamefont {Kim}}, \bibinfo
  {author} {\bibfnamefont {J.}~\bibnamefont {Paglione}}, \ and\ \bibinfo
  {author} {\bibfnamefont {N.~P.}\ \bibnamefont {Butch}},\ }\href@noop {}
  {\bibfield  {journal} {\bibinfo  {journal} {Science}\ }\textbf {\bibinfo
  {volume} {365}},\ \bibinfo {pages} {684} (\bibinfo {year}
  {2019}{\natexlab{a}})}\BibitemShut {NoStop}%
\bibitem [{\citenamefont {Aoki}\ \emph {et~al.}(2019)\citenamefont {Aoki},
  \citenamefont {Nakamura}, \citenamefont {Honda}, \citenamefont {Li},
  \citenamefont {Homma}, \citenamefont {Shimizu}, \citenamefont {Sato},
  \citenamefont {Knebel}, \citenamefont {Brison}, \citenamefont {Pourret},
  \citenamefont {Braithwaite}, \citenamefont {Lapertot}, \citenamefont {Niu},
  \citenamefont {Vališka}, \citenamefont {Harima},\ and\ \citenamefont
  {Flouquet}}]{Aoki2019b}%
  \BibitemOpen
  \bibfield  {author} {\bibinfo {author} {\bibfnamefont {D.}~\bibnamefont
  {Aoki}}, \bibinfo {author} {\bibfnamefont {A.}~\bibnamefont {Nakamura}},
  \bibinfo {author} {\bibfnamefont {F.}~\bibnamefont {Honda}}, \bibinfo
  {author} {\bibfnamefont {D.}~\bibnamefont {Li}}, \bibinfo {author}
  {\bibfnamefont {Y.}~\bibnamefont {Homma}}, \bibinfo {author} {\bibfnamefont
  {Y.}~\bibnamefont {Shimizu}}, \bibinfo {author} {\bibfnamefont {Y.~J.}\
  \bibnamefont {Sato}}, \bibinfo {author} {\bibfnamefont {G.}~\bibnamefont
  {Knebel}}, \bibinfo {author} {\bibfnamefont {J.-P.}\ \bibnamefont {Brison}},
  \bibinfo {author} {\bibfnamefont {A.}~\bibnamefont {Pourret}}, \bibinfo
  {author} {\bibfnamefont {D.}~\bibnamefont {Braithwaite}}, \bibinfo {author}
  {\bibfnamefont {G.}~\bibnamefont {Lapertot}}, \bibinfo {author}
  {\bibfnamefont {Q.}~\bibnamefont {Niu}}, \bibinfo {author} {\bibfnamefont
  {M.}~\bibnamefont {Vališka}}, \bibinfo {author} {\bibfnamefont
  {H.}~\bibnamefont {Harima}}, \ and\ \bibinfo {author} {\bibfnamefont
  {J.}~\bibnamefont {Flouquet}},\ }\href@noop {} {\bibfield  {journal}
  {\bibinfo  {journal} {J. Phys. Soc. Jpn.}\ }\textbf {\bibinfo {volume}
  {88}},\ \bibinfo {pages} {043702} (\bibinfo {year} {2019})}\BibitemShut
  {NoStop}%
\bibitem [{\citenamefont {Ishizuka}\ \emph {et~al.}(2019)\citenamefont
  {Ishizuka}, \citenamefont {Sumita}, \citenamefont {Daido},\ and\
  \citenamefont {Yanase}}]{Ishizuka2019}%
  \BibitemOpen
  \bibfield  {author} {\bibinfo {author} {\bibfnamefont {J.}~\bibnamefont
  {Ishizuka}}, \bibinfo {author} {\bibfnamefont {S.}~\bibnamefont {Sumita}},
  \bibinfo {author} {\bibfnamefont {A.}~\bibnamefont {Daido}}, \ and\ \bibinfo
  {author} {\bibfnamefont {Y.}~\bibnamefont {Yanase}},\ }\href@noop {}
  {\bibfield  {journal} {\bibinfo  {journal} {Phys. Rev. Lett.}\ }\textbf
  {\bibinfo {volume} {123}},\ \bibinfo {pages} {217001} (\bibinfo {year}
  {2019})}\BibitemShut {NoStop}%
\bibitem [{\citenamefont {Nakamine}\ \emph {et~al.}(2019)\citenamefont
  {Nakamine}, \citenamefont {Kitagawa}, \citenamefont {Ishida}, \citenamefont
  {Tokunaga}, \citenamefont {Sakai}, \citenamefont {Kambe}, \citenamefont
  {Nakamura}, \citenamefont {Shimizu}, \citenamefont {Homma}, \citenamefont
  {Li}, \citenamefont {Honda},\ and\ \citenamefont {Aoki}}]{Nakamine2019}%
  \BibitemOpen
  \bibfield  {author} {\bibinfo {author} {\bibfnamefont {G.}~\bibnamefont
  {Nakamine}}, \bibinfo {author} {\bibfnamefont {S.}~\bibnamefont {Kitagawa}},
  \bibinfo {author} {\bibfnamefont {K.}~\bibnamefont {Ishida}}, \bibinfo
  {author} {\bibfnamefont {Y.}~\bibnamefont {Tokunaga}}, \bibinfo {author}
  {\bibfnamefont {H.}~\bibnamefont {Sakai}}, \bibinfo {author} {\bibfnamefont
  {S.}~\bibnamefont {Kambe}}, \bibinfo {author} {\bibfnamefont
  {A.}~\bibnamefont {Nakamura}}, \bibinfo {author} {\bibfnamefont
  {Y.}~\bibnamefont {Shimizu}}, \bibinfo {author} {\bibfnamefont
  {Y.}~\bibnamefont {Homma}}, \bibinfo {author} {\bibfnamefont
  {D.}~\bibnamefont {Li}}, \bibinfo {author} {\bibfnamefont {F.}~\bibnamefont
  {Honda}}, \ and\ \bibinfo {author} {\bibfnamefont {D.}~\bibnamefont {Aoki}},\
  }\href@noop {} {\bibfield  {journal} {\bibinfo  {journal} {J. Phys. Soc.
  Jpn.}\ }\textbf {\bibinfo {volume} {88}},\ \bibinfo {pages} {113703}
  (\bibinfo {year} {2019})}\BibitemShut {NoStop}%
\bibitem [{\citenamefont {Metz}\ \emph {et~al.}(2019)\citenamefont {Metz},
  \citenamefont {Bae}, \citenamefont {Ran}, \citenamefont {Liu}, \citenamefont
  {Eo}, \citenamefont {Fuhrman}, \citenamefont {Agterberg}, \citenamefont
  {Anlage}, \citenamefont {Butch},\ and\ \citenamefont {Paglione}}]{Metz2019}%
  \BibitemOpen
  \bibfield  {author} {\bibinfo {author} {\bibfnamefont {T.}~\bibnamefont
  {Metz}}, \bibinfo {author} {\bibfnamefont {S.}~\bibnamefont {Bae}}, \bibinfo
  {author} {\bibfnamefont {S.}~\bibnamefont {Ran}}, \bibinfo {author}
  {\bibfnamefont {I.-L.}\ \bibnamefont {Liu}}, \bibinfo {author} {\bibfnamefont
  {Y.~S.}\ \bibnamefont {Eo}}, \bibinfo {author} {\bibfnamefont {W.~T.}\
  \bibnamefont {Fuhrman}}, \bibinfo {author} {\bibfnamefont {D.~F.}\
  \bibnamefont {Agterberg}}, \bibinfo {author} {\bibfnamefont {S.~M.}\
  \bibnamefont {Anlage}}, \bibinfo {author} {\bibfnamefont {N.~P.}\
  \bibnamefont {Butch}}, \ and\ \bibinfo {author} {\bibfnamefont
  {J.}~\bibnamefont {Paglione}},\ }\href@noop {} {\bibfield  {journal}
  {\bibinfo  {journal} {Phys. Rev. B}\ }\textbf {\bibinfo {volume} {100}},\
  \bibinfo {pages} {220504} (\bibinfo {year} {2019})}\BibitemShut {NoStop}%
\bibitem [{\citenamefont {Kittaka}\ \emph {et~al.}(2020)\citenamefont
  {Kittaka}, \citenamefont {Shimizu}, \citenamefont {Sakakibara}, \citenamefont
  {Nakamura}, \citenamefont {Li}, \citenamefont {Homma}, \citenamefont {Honda},
  \citenamefont {Aoki},\ and\ \citenamefont {Machida}}]{Kittaka2020}%
  \BibitemOpen
  \bibfield  {author} {\bibinfo {author} {\bibfnamefont {S.}~\bibnamefont
  {Kittaka}}, \bibinfo {author} {\bibfnamefont {Y.}~\bibnamefont {Shimizu}},
  \bibinfo {author} {\bibfnamefont {T.}~\bibnamefont {Sakakibara}}, \bibinfo
  {author} {\bibfnamefont {A.}~\bibnamefont {Nakamura}}, \bibinfo {author}
  {\bibfnamefont {D.}~\bibnamefont {Li}}, \bibinfo {author} {\bibfnamefont
  {Y.}~\bibnamefont {Homma}}, \bibinfo {author} {\bibfnamefont
  {F.}~\bibnamefont {Honda}}, \bibinfo {author} {\bibfnamefont
  {D.}~\bibnamefont {Aoki}}, \ and\ \bibinfo {author} {\bibfnamefont
  {K.}~\bibnamefont {Machida}},\ }\href@noop {} {\bibfield  {journal} {\bibinfo
   {journal} {Phys. Rev. Research}\ }\textbf {\bibinfo {volume} {2}},\ \bibinfo
  {pages} {032014} (\bibinfo {year} {2020})}\BibitemShut {NoStop}%
\bibitem [{\citenamefont {Jiao}\ \emph {et~al.}(2020)\citenamefont {Jiao},
  \citenamefont {Howard}, \citenamefont {Ran}, \citenamefont {Wang},
  \citenamefont {Rodriguez}, \citenamefont {Sigrist}, \citenamefont {Wang},
  \citenamefont {Butch},\ and\ \citenamefont {Madhavan}}]{Jiao2020}%
  \BibitemOpen
  \bibfield  {author} {\bibinfo {author} {\bibfnamefont {L.}~\bibnamefont
  {Jiao}}, \bibinfo {author} {\bibfnamefont {S.}~\bibnamefont {Howard}},
  \bibinfo {author} {\bibfnamefont {S.}~\bibnamefont {Ran}}, \bibinfo {author}
  {\bibfnamefont {Z.}~\bibnamefont {Wang}}, \bibinfo {author} {\bibfnamefont
  {J.~O.}\ \bibnamefont {Rodriguez}}, \bibinfo {author} {\bibfnamefont
  {M.}~\bibnamefont {Sigrist}}, \bibinfo {author} {\bibfnamefont
  {Z.}~\bibnamefont {Wang}}, \bibinfo {author} {\bibfnamefont {N.~P.}\
  \bibnamefont {Butch}}, \ and\ \bibinfo {author} {\bibfnamefont
  {V.}~\bibnamefont {Madhavan}},\ }\href@noop {} {\bibfield  {journal}
  {\bibinfo  {journal} {Nature}\ }\textbf {\bibinfo {volume} {579}},\ \bibinfo
  {pages} {523} (\bibinfo {year} {2020})}\BibitemShut {NoStop}%
\bibitem [{\citenamefont {Shishidou}\ \emph {et~al.}(2021)\citenamefont
  {Shishidou}, \citenamefont {Suh}, \citenamefont {Brydon}, \citenamefont
  {Weinert},\ and\ \citenamefont {Agterberg}}]{Shishidou2021}%
  \BibitemOpen
  \bibfield  {author} {\bibinfo {author} {\bibfnamefont {T.}~\bibnamefont
  {Shishidou}}, \bibinfo {author} {\bibfnamefont {H.~G.}\ \bibnamefont {Suh}},
  \bibinfo {author} {\bibfnamefont {P.~M.~R.}\ \bibnamefont {Brydon}}, \bibinfo
  {author} {\bibfnamefont {M.}~\bibnamefont {Weinert}}, \ and\ \bibinfo
  {author} {\bibfnamefont {D.~F.}\ \bibnamefont {Agterberg}},\ }\href@noop {}
  {\bibfield  {journal} {\bibinfo  {journal} {Phys. Rev. B}\ }\textbf {\bibinfo
  {volume} {103}},\ \bibinfo {pages} {104504} (\bibinfo {year}
  {2021})}\BibitemShut {NoStop}%
\bibitem [{\citenamefont {Hayes}\ \emph {et~al.}()\citenamefont {Hayes},
  \citenamefont {Wei}, \citenamefont {Metz}, \citenamefont {Zhang},
  \citenamefont {Eo}, \citenamefont {Ran}, \citenamefont {Saha}, \citenamefont
  {Collini}, \citenamefont {Butch}, \citenamefont {Agterberg}, \citenamefont
  {Kapitulnik},\ and\ \citenamefont {Paglione}}]{Hayes2021}%
  \BibitemOpen
  \bibfield  {author} {\bibinfo {author} {\bibfnamefont {I.~M.}\ \bibnamefont
  {Hayes}}, \bibinfo {author} {\bibfnamefont {D.~S.}\ \bibnamefont {Wei}},
  \bibinfo {author} {\bibfnamefont {T.}~\bibnamefont {Metz}}, \bibinfo {author}
  {\bibfnamefont {J.}~\bibnamefont {Zhang}}, \bibinfo {author} {\bibfnamefont
  {Y.~S.}\ \bibnamefont {Eo}}, \bibinfo {author} {\bibfnamefont
  {S.}~\bibnamefont {Ran}}, \bibinfo {author} {\bibfnamefont {S.~R.}\
  \bibnamefont {Saha}}, \bibinfo {author} {\bibfnamefont {J.}~\bibnamefont
  {Collini}}, \bibinfo {author} {\bibfnamefont {N.~P.}\ \bibnamefont {Butch}},
  \bibinfo {author} {\bibfnamefont {D.~F.}\ \bibnamefont {Agterberg}}, \bibinfo
  {author} {\bibfnamefont {A.}~\bibnamefont {Kapitulnik}}, \ and\ \bibinfo
  {author} {\bibfnamefont {J.}~\bibnamefont {Paglione}},\ }\href@noop {}
  {\bibinfo  {journal} {arXiv:2002.02539}\ }\BibitemShut {NoStop}%
\bibitem [{\citenamefont {Bae}\ \emph {et~al.}(2021)\citenamefont {Bae},
  \citenamefont {Kim}, \citenamefont {Eo}, \citenamefont {Ran}, \citenamefont
  {lin Liu}, \citenamefont {Fuhrman}, \citenamefont {Paglione}, \citenamefont
  {Butch},\ and\ \citenamefont {Anlage}}]{Bae2021}%
  \BibitemOpen
\bibfield  {journal} {  }\bibfield  {author} {\bibinfo {author} {\bibfnamefont
  {S.}~\bibnamefont {Bae}}, \bibinfo {author} {\bibfnamefont {H.}~\bibnamefont
  {Kim}}, \bibinfo {author} {\bibfnamefont {Y.~S.}\ \bibnamefont {Eo}},
  \bibinfo {author} {\bibfnamefont {S.}~\bibnamefont {Ran}}, \bibinfo {author}
  {\bibfnamefont {I.}~\bibnamefont {lin Liu}}, \bibinfo {author} {\bibfnamefont
  {W.~T.}\ \bibnamefont {Fuhrman}}, \bibinfo {author} {\bibfnamefont
  {J.}~\bibnamefont {Paglione}}, \bibinfo {author} {\bibfnamefont {N.~P.}\
  \bibnamefont {Butch}}, \ and\ \bibinfo {author} {\bibfnamefont {S.~M.}\
  \bibnamefont {Anlage}},\ }\href@noop {} {\bibfield  {journal} {\bibinfo
  {journal} {Nat. Commun.}\ }\textbf {\bibinfo {volume} {12}},\ \bibinfo
  {pages} {2644} (\bibinfo {year} {2021})}\BibitemShut {NoStop}%
\bibitem [{\citenamefont {Sato}\ and\ \citenamefont {Ando}(2017)}]{Sato2017}%
  \BibitemOpen
  \bibfield  {author} {\bibinfo {author} {\bibfnamefont {M.}~\bibnamefont
  {Sato}}\ and\ \bibinfo {author} {\bibfnamefont {Y.}~\bibnamefont {Ando}},\
  }\href@noop {} {\bibfield  {journal} {\bibinfo  {journal} {Rep. Prog. Phys.}\
  }\textbf {\bibinfo {volume} {80}},\ \bibinfo {pages} {076501} (\bibinfo
  {year} {2017})}\BibitemShut {NoStop}%
\bibitem [{\citenamefont {Shick}\ \emph {et~al.}(2021)\citenamefont {Shick},
  \citenamefont {Fujimori},\ and\ \citenamefont {Pickett}}]{Shick2021}%
  \BibitemOpen
  \bibfield  {author} {\bibinfo {author} {\bibfnamefont {A.~B.}\ \bibnamefont
  {Shick}}, \bibinfo {author} {\bibfnamefont {S.-i.}\ \bibnamefont {Fujimori}},
  \ and\ \bibinfo {author} {\bibfnamefont {W.~E.}\ \bibnamefont {Pickett}},\
  }\href@noop {} {\bibfield  {journal} {\bibinfo  {journal} {Phys. Rev. B}\
  }\textbf {\bibinfo {volume} {103}},\ \bibinfo {pages} {125136} (\bibinfo
  {year} {2021})}\BibitemShut {NoStop}%
\bibitem [{\citenamefont {Ishizuka}\ and\ \citenamefont
  {Yanase}(2021)}]{Ishizuka2021}%
  \BibitemOpen
  \bibfield  {author} {\bibinfo {author} {\bibfnamefont {J.}~\bibnamefont
  {Ishizuka}}\ and\ \bibinfo {author} {\bibfnamefont {Y.}~\bibnamefont
  {Yanase}},\ }\href@noop {} {\bibfield  {journal} {\bibinfo  {journal} {Phys.
  Rev. B}\ }\textbf {\bibinfo {volume} {103}},\ \bibinfo {pages} {094504}
  (\bibinfo {year} {2021})}\BibitemShut {NoStop}%
\bibitem [{\citenamefont {Miao}\ \emph {et~al.}(2020)\citenamefont {Miao},
  \citenamefont {Liu}, \citenamefont {Xu}, \citenamefont {Kotta}, \citenamefont
  {Kang}, \citenamefont {Ran}, \citenamefont {Paglione}, \citenamefont
  {Kotliar}, \citenamefont {Butch}, \citenamefont {Denlinger},\ and\
  \citenamefont {Wray}}]{Miao2020}%
  \BibitemOpen
  \bibfield  {author} {\bibinfo {author} {\bibfnamefont {L.}~\bibnamefont
  {Miao}}, \bibinfo {author} {\bibfnamefont {S.}~\bibnamefont {Liu}}, \bibinfo
  {author} {\bibfnamefont {Y.}~\bibnamefont {Xu}}, \bibinfo {author}
  {\bibfnamefont {E.~C.}\ \bibnamefont {Kotta}}, \bibinfo {author}
  {\bibfnamefont {C.-J.}\ \bibnamefont {Kang}}, \bibinfo {author}
  {\bibfnamefont {S.}~\bibnamefont {Ran}}, \bibinfo {author} {\bibfnamefont
  {J.}~\bibnamefont {Paglione}}, \bibinfo {author} {\bibfnamefont
  {G.}~\bibnamefont {Kotliar}}, \bibinfo {author} {\bibfnamefont {N.~P.}\
  \bibnamefont {Butch}}, \bibinfo {author} {\bibfnamefont {J.~D.}\ \bibnamefont
  {Denlinger}}, \ and\ \bibinfo {author} {\bibfnamefont {L.~A.}\ \bibnamefont
  {Wray}},\ }\href@noop {} {\bibfield  {journal} {\bibinfo  {journal} {Phys.
  Rev. Lett.}\ }\textbf {\bibinfo {volume} {124}},\ \bibinfo {pages} {076401}
  (\bibinfo {year} {2020})}\BibitemShut {NoStop}%
\bibitem [{\citenamefont {Xu}\ \emph {et~al.}(2019)\citenamefont {Xu},
  \citenamefont {Sheng},\ and\ \citenamefont {Yang}}]{Xu2019}%
  \BibitemOpen
  \bibfield  {author} {\bibinfo {author} {\bibfnamefont {Y.}~\bibnamefont
  {Xu}}, \bibinfo {author} {\bibfnamefont {Y.}~\bibnamefont {Sheng}}, \ and\
  \bibinfo {author} {\bibfnamefont {Y.-f.}\ \bibnamefont {Yang}},\ }\href@noop
  {} {\bibfield  {journal} {\bibinfo  {journal} {Phys. Rev. Lett.}\ }\textbf
  {\bibinfo {volume} {123}},\ \bibinfo {pages} {217002} (\bibinfo {year}
  {2019})}\BibitemShut {NoStop}%
\bibitem [{\citenamefont {Machida}(2020)}]{Machida2020}%
  \BibitemOpen
  \bibfield  {author} {\bibinfo {author} {\bibfnamefont {K.}~\bibnamefont
  {Machida}},\ }\href@noop {} {\bibfield  {journal} {\bibinfo  {journal}
  {Journal of the Physical Society of Japan}\ }\textbf {\bibinfo {volume}
  {89}},\ \bibinfo {pages} {033702} (\bibinfo {year} {2020})}\BibitemShut
  {NoStop}%
\bibitem [{\citenamefont {Braithwaite}\ \emph {et~al.}(2019)\citenamefont
  {Braithwaite}, \citenamefont {Vali\v{s}ka}, \citenamefont {Knebel},
  \citenamefont {Lapertot}, \citenamefont {Brison}, \citenamefont {Pourret},
  \citenamefont {Zhitomirsky}, \citenamefont {Flouquet}, \citenamefont
  {Honda},\ and\ \citenamefont {Aoki}}]{Braithwaite2019}%
  \BibitemOpen
  \bibfield  {author} {\bibinfo {author} {\bibfnamefont {D.}~\bibnamefont
  {Braithwaite}}, \bibinfo {author} {\bibfnamefont {M.}~\bibnamefont
  {Vali\v{s}ka}}, \bibinfo {author} {\bibfnamefont {G.}~\bibnamefont {Knebel}},
  \bibinfo {author} {\bibfnamefont {G.}~\bibnamefont {Lapertot}}, \bibinfo
  {author} {\bibfnamefont {J.-P.}\ \bibnamefont {Brison}}, \bibinfo {author}
  {\bibfnamefont {A.}~\bibnamefont {Pourret}}, \bibinfo {author} {\bibfnamefont
  {M.~E.}\ \bibnamefont {Zhitomirsky}}, \bibinfo {author} {\bibfnamefont
  {J.}~\bibnamefont {Flouquet}}, \bibinfo {author} {\bibfnamefont
  {F.}~\bibnamefont {Honda}}, \ and\ \bibinfo {author} {\bibfnamefont
  {D.}~\bibnamefont {Aoki}},\ }\href@noop {} {\bibfield  {journal} {\bibinfo
  {journal} {Commun. Phys.}\ }\textbf {\bibinfo {volume} {2}},\ \bibinfo
  {pages} {147} (\bibinfo {year} {2019})}\BibitemShut {NoStop}%
\bibitem [{\citenamefont {Knafo}\ \emph {et~al.}(2019)\citenamefont {Knafo},
  \citenamefont {Vali\v{s}ka}, \citenamefont {Braithwaite}, \citenamefont
  {Lapertot}, \citenamefont {Knebel}, \citenamefont {Pourret}, \citenamefont
  {Brison}, \citenamefont {Flouquet},\ and\ \citenamefont {Aoki}}]{Knafo2019}%
  \BibitemOpen
  \bibfield  {author} {\bibinfo {author} {\bibfnamefont {W.}~\bibnamefont
  {Knafo}}, \bibinfo {author} {\bibfnamefont {M.}~\bibnamefont {Vali\v{s}ka}},
  \bibinfo {author} {\bibfnamefont {D.}~\bibnamefont {Braithwaite}}, \bibinfo
  {author} {\bibfnamefont {G.}~\bibnamefont {Lapertot}}, \bibinfo {author}
  {\bibfnamefont {G.}~\bibnamefont {Knebel}}, \bibinfo {author} {\bibfnamefont
  {A.}~\bibnamefont {Pourret}}, \bibinfo {author} {\bibfnamefont {J.-P.}\
  \bibnamefont {Brison}}, \bibinfo {author} {\bibfnamefont {J.}~\bibnamefont
  {Flouquet}}, \ and\ \bibinfo {author} {\bibfnamefont {D.}~\bibnamefont
  {Aoki}},\ }\href@noop {} {\bibfield  {journal} {\bibinfo  {journal} {J. Phys.
  Soc. Jpn.}\ }\textbf {\bibinfo {volume} {88}},\ \bibinfo {pages} {063705}
  (\bibinfo {year} {2019})}\BibitemShut {NoStop}%
\bibitem [{\citenamefont {Miyake}\ \emph {et~al.}(2019)\citenamefont {Miyake},
  \citenamefont {Shimizu}, \citenamefont {Sato}, \citenamefont {Li},
  \citenamefont {Nakamura}, \citenamefont {Homma}, \citenamefont {Honda},
  \citenamefont {Flouquet}, \citenamefont {Tokunaga},\ and\ \citenamefont
  {Aoki}}]{Miyake2019}%
  \BibitemOpen
  \bibfield  {author} {\bibinfo {author} {\bibfnamefont {A.}~\bibnamefont
  {Miyake}}, \bibinfo {author} {\bibfnamefont {Y.}~\bibnamefont {Shimizu}},
  \bibinfo {author} {\bibfnamefont {Y.~J.}\ \bibnamefont {Sato}}, \bibinfo
  {author} {\bibfnamefont {D.}~\bibnamefont {Li}}, \bibinfo {author}
  {\bibfnamefont {A.}~\bibnamefont {Nakamura}}, \bibinfo {author}
  {\bibfnamefont {Y.}~\bibnamefont {Homma}}, \bibinfo {author} {\bibfnamefont
  {F.}~\bibnamefont {Honda}}, \bibinfo {author} {\bibfnamefont
  {J.}~\bibnamefont {Flouquet}}, \bibinfo {author} {\bibfnamefont
  {M.}~\bibnamefont {Tokunaga}}, \ and\ \bibinfo {author} {\bibfnamefont
  {D.}~\bibnamefont {Aoki}},\ }\href@noop {} {\bibfield  {journal} {\bibinfo
  {journal} {J. Phys. Soc. Jpn.}\ }\textbf {\bibinfo {volume} {88}},\ \bibinfo
  {pages} {063706} (\bibinfo {year} {2019})}\BibitemShut {NoStop}%
\bibitem [{\citenamefont {Knebel}\ \emph {et~al.}(2019)\citenamefont {Knebel},
  \citenamefont {Knafo}, \citenamefont {Pourret}, \citenamefont {Niu},
  \citenamefont {Vali\v{s}ka}, \citenamefont {Braithwaite}, \citenamefont
  {Lapertot}, \citenamefont {Nardone}, \citenamefont {Zitouni}, \citenamefont
  {Mishra}, \citenamefont {Sheikin}, \citenamefont {Seyfarth}, \citenamefont
  {Brison}, \citenamefont {Aoki},\ and\ \citenamefont {Flouquet}}]{Knebel2019}%
  \BibitemOpen
  \bibfield  {author} {\bibinfo {author} {\bibfnamefont {G.}~\bibnamefont
  {Knebel}}, \bibinfo {author} {\bibfnamefont {W.}~\bibnamefont {Knafo}},
  \bibinfo {author} {\bibfnamefont {A.}~\bibnamefont {Pourret}}, \bibinfo
  {author} {\bibfnamefont {Q.}~\bibnamefont {Niu}}, \bibinfo {author}
  {\bibfnamefont {M.}~\bibnamefont {Vali\v{s}ka}}, \bibinfo {author}
  {\bibfnamefont {D.}~\bibnamefont {Braithwaite}}, \bibinfo {author}
  {\bibfnamefont {G.}~\bibnamefont {Lapertot}}, \bibinfo {author}
  {\bibfnamefont {M.}~\bibnamefont {Nardone}}, \bibinfo {author} {\bibfnamefont
  {A.}~\bibnamefont {Zitouni}}, \bibinfo {author} {\bibfnamefont
  {S.}~\bibnamefont {Mishra}}, \bibinfo {author} {\bibfnamefont
  {I.}~\bibnamefont {Sheikin}}, \bibinfo {author} {\bibfnamefont
  {G.}~\bibnamefont {Seyfarth}}, \bibinfo {author} {\bibfnamefont {J.-P.}\
  \bibnamefont {Brison}}, \bibinfo {author} {\bibfnamefont {D.}~\bibnamefont
  {Aoki}}, \ and\ \bibinfo {author} {\bibfnamefont {J.}~\bibnamefont
  {Flouquet}},\ }\href@noop {} {\bibfield  {journal} {\bibinfo  {journal} {J.
  Phys. Soc. Jpn.}\ }\textbf {\bibinfo {volume} {88}},\ \bibinfo {pages}
  {063707} (\bibinfo {year} {2019})}\BibitemShut {NoStop}%
\bibitem [{\citenamefont {Ran}\ \emph {et~al.}(2019{\natexlab{b}})\citenamefont
  {Ran}, \citenamefont {Liu}, \citenamefont {Eo}, \citenamefont {Campbell},
  \citenamefont {Neves}, \citenamefont {Fuhrman}, \citenamefont {Saha},
  \citenamefont {Eckberg}, \citenamefont {Kim}, \citenamefont {Paglione},
  \citenamefont {Graf}, \citenamefont {Singleton},\ and\ \citenamefont
  {Butch}}]{Ran2019b}%
  \BibitemOpen
  \bibfield  {author} {\bibinfo {author} {\bibfnamefont {S.}~\bibnamefont
  {Ran}}, \bibinfo {author} {\bibfnamefont {I.-L.}\ \bibnamefont {Liu}},
  \bibinfo {author} {\bibfnamefont {Y.~S.}\ \bibnamefont {Eo}}, \bibinfo
  {author} {\bibfnamefont {D.~J.}\ \bibnamefont {Campbell}}, \bibinfo {author}
  {\bibfnamefont {P.}~\bibnamefont {Neves}}, \bibinfo {author} {\bibfnamefont
  {W.~T.}\ \bibnamefont {Fuhrman}}, \bibinfo {author} {\bibfnamefont {S.~R.}\
  \bibnamefont {Saha}}, \bibinfo {author} {\bibfnamefont {C.}~\bibnamefont
  {Eckberg}}, \bibinfo {author} {\bibfnamefont {H.}~\bibnamefont {Kim}},
  \bibinfo {author} {\bibfnamefont {J.}~\bibnamefont {Paglione}}, \bibinfo
  {author} {\bibfnamefont {D.}~\bibnamefont {Graf}}, \bibinfo {author}
  {\bibfnamefont {J.}~\bibnamefont {Singleton}}, \ and\ \bibinfo {author}
  {\bibfnamefont {N.~P.}\ \bibnamefont {Butch}},\ }\href@noop {} {\bibfield
  {journal} {\bibinfo  {journal} {Nat. Phys.}\ }\textbf {\bibinfo {volume}
  {15}},\ \bibinfo {pages} {1250} (\bibinfo {year}
  {2019}{\natexlab{b}})}\BibitemShut {NoStop}%
\bibitem [{\citenamefont {Aoki}\ \emph {et~al.}(2020)\citenamefont {Aoki},
  \citenamefont {Honda}, \citenamefont {Knebel}, \citenamefont {Braithwaite},
  \citenamefont {Nakamura}, \citenamefont {Li}, \citenamefont {Homma},
  \citenamefont {Shimizu}, \citenamefont {Sato}, \citenamefont {Brison},\ and\
  \citenamefont {Flouquet}}]{Aoki2020}%
  \BibitemOpen
  \bibfield  {author} {\bibinfo {author} {\bibfnamefont {D.}~\bibnamefont
  {Aoki}}, \bibinfo {author} {\bibfnamefont {F.}~\bibnamefont {Honda}},
  \bibinfo {author} {\bibfnamefont {G.}~\bibnamefont {Knebel}}, \bibinfo
  {author} {\bibfnamefont {D.}~\bibnamefont {Braithwaite}}, \bibinfo {author}
  {\bibfnamefont {A.}~\bibnamefont {Nakamura}}, \bibinfo {author}
  {\bibfnamefont {D.}~\bibnamefont {Li}}, \bibinfo {author} {\bibfnamefont
  {Y.}~\bibnamefont {Homma}}, \bibinfo {author} {\bibfnamefont
  {Y.}~\bibnamefont {Shimizu}}, \bibinfo {author} {\bibfnamefont {Y.~J.}\
  \bibnamefont {Sato}}, \bibinfo {author} {\bibfnamefont {J.-P.}\ \bibnamefont
  {Brison}}, \ and\ \bibinfo {author} {\bibfnamefont {J.}~\bibnamefont
  {Flouquet}},\ }\href@noop {} {\bibfield  {journal} {\bibinfo  {journal} {J.
  Phys. Soc. Jpn.}\ }\textbf {\bibinfo {volume} {89}},\ \bibinfo {pages}
  {053705} (\bibinfo {year} {2020})}\BibitemShut {NoStop}%
\bibitem [{\citenamefont {Ran}\ \emph {et~al.}(2020)\citenamefont {Ran},
  \citenamefont {Kim}, \citenamefont {Liu}, \citenamefont {Saha}, \citenamefont
  {Hayes}, \citenamefont {Metz}, \citenamefont {Eo}, \citenamefont {Paglione},\
  and\ \citenamefont {Butch}}]{Ran2020}%
  \BibitemOpen
  \bibfield  {author} {\bibinfo {author} {\bibfnamefont {S.}~\bibnamefont
  {Ran}}, \bibinfo {author} {\bibfnamefont {H.}~\bibnamefont {Kim}}, \bibinfo
  {author} {\bibfnamefont {I.-L.}\ \bibnamefont {Liu}}, \bibinfo {author}
  {\bibfnamefont {S.~R.}\ \bibnamefont {Saha}}, \bibinfo {author}
  {\bibfnamefont {I.}~\bibnamefont {Hayes}}, \bibinfo {author} {\bibfnamefont
  {T.}~\bibnamefont {Metz}}, \bibinfo {author} {\bibfnamefont {Y.~S.}\
  \bibnamefont {Eo}}, \bibinfo {author} {\bibfnamefont {J.}~\bibnamefont
  {Paglione}}, \ and\ \bibinfo {author} {\bibfnamefont {N.~P.}\ \bibnamefont
  {Butch}},\ }\href@noop {} {\bibfield  {journal} {\bibinfo  {journal} {Phys.
  Rev. B}\ }\textbf {\bibinfo {volume} {101}},\ \bibinfo {pages} {140503}
  (\bibinfo {year} {2020})}\BibitemShut {NoStop}%
\bibitem [{\citenamefont {Lin}\ \emph {et~al.}(2020)\citenamefont {Lin},
  \citenamefont {Campbell}, \citenamefont {Ran}, \citenamefont {Liu},
  \citenamefont {Kim}, \citenamefont {Nevidomskyy}, \citenamefont {Graf},
  \citenamefont {Butch},\ and\ \citenamefont {Paglione}}]{Lin2020}%
  \BibitemOpen
  \bibfield  {author} {\bibinfo {author} {\bibfnamefont {W.~C.}\ \bibnamefont
  {Lin}}, \bibinfo {author} {\bibfnamefont {D.~J.}\ \bibnamefont {Campbell}},
  \bibinfo {author} {\bibfnamefont {S.}~\bibnamefont {Ran}}, \bibinfo {author}
  {\bibfnamefont {I.-L.}\ \bibnamefont {Liu}}, \bibinfo {author} {\bibfnamefont
  {H.}~\bibnamefont {Kim}}, \bibinfo {author} {\bibfnamefont {A.~H.}\
  \bibnamefont {Nevidomskyy}}, \bibinfo {author} {\bibfnamefont
  {D.}~\bibnamefont {Graf}}, \bibinfo {author} {\bibfnamefont {N.~P.}\
  \bibnamefont {Butch}}, \ and\ \bibinfo {author} {\bibfnamefont
  {J.}~\bibnamefont {Paglione}},\ }\href@noop {} {\bibfield  {journal}
  {\bibinfo  {journal} {npj Quantum Mater.}\ }\textbf {\bibinfo {volume} {5}},\
  \bibinfo {pages} {68} (\bibinfo {year} {2020})}\BibitemShut {NoStop}%
\bibitem [{\citenamefont {Knebel}\ \emph {et~al.}(2020)\citenamefont {Knebel},
  \citenamefont {Kimata}, \citenamefont {Vali\v{s}ka}, \citenamefont {Honda},
  \citenamefont {Li}, \citenamefont {Braithwaite}, \citenamefont {Lapertot},
  \citenamefont {Knafo}, \citenamefont {Pourret}, \citenamefont {Sato},
  \citenamefont {Shimizu}, \citenamefont {Kihara}, \citenamefont {Brison},
  \citenamefont {Flouquet},\ and\ \citenamefont {Aoki}}]{Knebel2020}%
  \BibitemOpen
  \bibfield  {author} {\bibinfo {author} {\bibfnamefont {G.}~\bibnamefont
  {Knebel}}, \bibinfo {author} {\bibfnamefont {M.}~\bibnamefont {Kimata}},
  \bibinfo {author} {\bibfnamefont {M.}~\bibnamefont {Vali\v{s}ka}}, \bibinfo
  {author} {\bibfnamefont {F.}~\bibnamefont {Honda}}, \bibinfo {author}
  {\bibfnamefont {D.}~\bibnamefont {Li}}, \bibinfo {author} {\bibfnamefont
  {D.}~\bibnamefont {Braithwaite}}, \bibinfo {author} {\bibfnamefont
  {G.}~\bibnamefont {Lapertot}}, \bibinfo {author} {\bibfnamefont
  {W.}~\bibnamefont {Knafo}}, \bibinfo {author} {\bibfnamefont
  {A.}~\bibnamefont {Pourret}}, \bibinfo {author} {\bibfnamefont {Y.~J.}\
  \bibnamefont {Sato}}, \bibinfo {author} {\bibfnamefont {Y.}~\bibnamefont
  {Shimizu}}, \bibinfo {author} {\bibfnamefont {T.}~\bibnamefont {Kihara}},
  \bibinfo {author} {\bibfnamefont {J.-P.}\ \bibnamefont {Brison}}, \bibinfo
  {author} {\bibfnamefont {J.}~\bibnamefont {Flouquet}}, \ and\ \bibinfo
  {author} {\bibfnamefont {D.}~\bibnamefont {Aoki}},\ }\href@noop {} {\bibfield
   {journal} {\bibinfo  {journal} {Journal of the Physical Society of Japan}\
  }\textbf {\bibinfo {volume} {89}},\ \bibinfo {pages} {053707} (\bibinfo
  {year} {2020})}\BibitemShut {NoStop}%
\bibitem [{\citenamefont {Knafo}\ \emph {et~al.}(2021)\citenamefont {Knafo},
  \citenamefont {Nardone}, \citenamefont {Vali\v{s}ka}, \citenamefont
  {Zitouni}, \citenamefont {Lapertot}, \citenamefont {Aoki}, \citenamefont
  {Knebel},\ and\ \citenamefont {Braithwaite}}]{Knafo2021}%
  \BibitemOpen
  \bibfield  {author} {\bibinfo {author} {\bibfnamefont {W.}~\bibnamefont
  {Knafo}}, \bibinfo {author} {\bibfnamefont {M.}~\bibnamefont {Nardone}},
  \bibinfo {author} {\bibfnamefont {M.}~\bibnamefont {Vali\v{s}ka}}, \bibinfo
  {author} {\bibfnamefont {A.}~\bibnamefont {Zitouni}}, \bibinfo {author}
  {\bibfnamefont {G.}~\bibnamefont {Lapertot}}, \bibinfo {author}
  {\bibfnamefont {D.}~\bibnamefont {Aoki}}, \bibinfo {author} {\bibfnamefont
  {G.}~\bibnamefont {Knebel}}, \ and\ \bibinfo {author} {\bibfnamefont
  {D.}~\bibnamefont {Braithwaite}},\ }\href@noop {} {\bibfield  {journal}
  {\bibinfo  {journal} {Commun. Phys.}\ }\textbf {\bibinfo {volume} {4}},\
  \bibinfo {pages} {40} (\bibinfo {year} {2021})}\BibitemShut {NoStop}%
\bibitem [{\citenamefont {Aoki}\ \emph {et~al.}(2021)\citenamefont {Aoki},
  \citenamefont {Kimata}, \citenamefont {Sato}, \citenamefont {Knebel},
  \citenamefont {Honda}, \citenamefont {Nakamura}, \citenamefont {Li},
  \citenamefont {Homma}, \citenamefont {Shimizu}, \citenamefont {Knafo},
  \citenamefont {Braithwaite}, \citenamefont {Vali\v{s}ka}, \citenamefont
  {Pourret}, \citenamefont {Brison},\ and\ \citenamefont
  {Flouquet}}]{Aoki2021}%
  \BibitemOpen
  \bibfield  {author} {\bibinfo {author} {\bibfnamefont {D.}~\bibnamefont
  {Aoki}}, \bibinfo {author} {\bibfnamefont {M.}~\bibnamefont {Kimata}},
  \bibinfo {author} {\bibfnamefont {Y.~J.}\ \bibnamefont {Sato}}, \bibinfo
  {author} {\bibfnamefont {G.}~\bibnamefont {Knebel}}, \bibinfo {author}
  {\bibfnamefont {F.}~\bibnamefont {Honda}}, \bibinfo {author} {\bibfnamefont
  {A.}~\bibnamefont {Nakamura}}, \bibinfo {author} {\bibfnamefont
  {D.}~\bibnamefont {Li}}, \bibinfo {author} {\bibfnamefont {Y.}~\bibnamefont
  {Homma}}, \bibinfo {author} {\bibfnamefont {Y.}~\bibnamefont {Shimizu}},
  \bibinfo {author} {\bibfnamefont {W.}~\bibnamefont {Knafo}}, \bibinfo
  {author} {\bibfnamefont {D.}~\bibnamefont {Braithwaite}}, \bibinfo {author}
  {\bibfnamefont {M.}~\bibnamefont {Vali\v{s}ka}}, \bibinfo {author}
  {\bibfnamefont {A.}~\bibnamefont {Pourret}}, \bibinfo {author} {\bibfnamefont
  {J.-P.}\ \bibnamefont {Brison}}, \ and\ \bibinfo {author} {\bibfnamefont
  {J.}~\bibnamefont {Flouquet}},\ }\href@noop {} {\bibfield  {journal}
  {\bibinfo  {journal} {J. Phys. Soc. Jpn.}\ }\textbf {\bibinfo {volume}
  {90}},\ \bibinfo {pages} {074705} (\bibinfo {year} {2021})}\BibitemShut
  {NoStop}%
\bibitem [{\citenamefont {Tokunaga}\ \emph {et~al.}(2019)\citenamefont
  {Tokunaga}, \citenamefont {Sakai}, \citenamefont {Kambe}, \citenamefont
  {Hattori}, \citenamefont {Higa}, \citenamefont {Nakamine}, \citenamefont
  {Kitagawa}, \citenamefont {Ishida}, \citenamefont {Nakamura}, \citenamefont
  {Shimizu}, \citenamefont {Homma}, \citenamefont {Li}, \citenamefont {Honda},\
  and\ \citenamefont {Aoki}}]{Tokunaga2019}%
  \BibitemOpen
  \bibfield  {author} {\bibinfo {author} {\bibfnamefont {Y.}~\bibnamefont
  {Tokunaga}}, \bibinfo {author} {\bibfnamefont {H.}~\bibnamefont {Sakai}},
  \bibinfo {author} {\bibfnamefont {S.}~\bibnamefont {Kambe}}, \bibinfo
  {author} {\bibfnamefont {T.}~\bibnamefont {Hattori}}, \bibinfo {author}
  {\bibfnamefont {N.}~\bibnamefont {Higa}}, \bibinfo {author} {\bibfnamefont
  {G.}~\bibnamefont {Nakamine}}, \bibinfo {author} {\bibfnamefont
  {S.}~\bibnamefont {Kitagawa}}, \bibinfo {author} {\bibfnamefont
  {K.}~\bibnamefont {Ishida}}, \bibinfo {author} {\bibfnamefont
  {A.}~\bibnamefont {Nakamura}}, \bibinfo {author} {\bibfnamefont
  {Y.}~\bibnamefont {Shimizu}}, \bibinfo {author} {\bibfnamefont
  {Y.}~\bibnamefont {Homma}}, \bibinfo {author} {\bibfnamefont
  {D.}~\bibnamefont {Li}}, \bibinfo {author} {\bibfnamefont {F.}~\bibnamefont
  {Honda}}, \ and\ \bibinfo {author} {\bibfnamefont {D.}~\bibnamefont {Aoki}},\
  }\href@noop {} {\bibfield  {journal} {\bibinfo  {journal} {J. Phys. Soc.
  Jpn.}\ }\textbf {\bibinfo {volume} {88}},\ \bibinfo {pages} {073701}
  (\bibinfo {year} {2019})}\BibitemShut {NoStop}%
\bibitem [{\citenamefont {Sundar}\ \emph {et~al.}(2019)\citenamefont {Sundar},
  \citenamefont {Gheidi}, \citenamefont {Akintola}, \citenamefont {C\^ot\'e},
  \citenamefont {Dunsiger}, \citenamefont {Ran}, \citenamefont {Butch},
  \citenamefont {Saha}, \citenamefont {Paglione},\ and\ \citenamefont
  {Sonier}}]{Sundar2019}%
  \BibitemOpen
  \bibfield  {author} {\bibinfo {author} {\bibfnamefont {S.}~\bibnamefont
  {Sundar}}, \bibinfo {author} {\bibfnamefont {S.}~\bibnamefont {Gheidi}},
  \bibinfo {author} {\bibfnamefont {K.}~\bibnamefont {Akintola}}, \bibinfo
  {author} {\bibfnamefont {A.~M.}\ \bibnamefont {C\^ot\'e}}, \bibinfo {author}
  {\bibfnamefont {S.~R.}\ \bibnamefont {Dunsiger}}, \bibinfo {author}
  {\bibfnamefont {S.}~\bibnamefont {Ran}}, \bibinfo {author} {\bibfnamefont
  {N.~P.}\ \bibnamefont {Butch}}, \bibinfo {author} {\bibfnamefont {S.~R.}\
  \bibnamefont {Saha}}, \bibinfo {author} {\bibfnamefont {J.}~\bibnamefont
  {Paglione}}, \ and\ \bibinfo {author} {\bibfnamefont {J.~E.}\ \bibnamefont
  {Sonier}},\ }\href@noop {} {\bibfield  {journal} {\bibinfo  {journal} {Phys.
  Rev. B}\ }\textbf {\bibinfo {volume} {100}},\ \bibinfo {pages} {140502}
  (\bibinfo {year} {2019})}\BibitemShut {NoStop}%
\bibitem [{\citenamefont {Thomas}\ \emph {et~al.}(2020)\citenamefont {Thomas},
  \citenamefont {Santos}, \citenamefont {Christensen}, \citenamefont {Asaba},
  \citenamefont {Ronning}, \citenamefont {Thompson}, \citenamefont {Bauer},
  \citenamefont {Fernandes}, \citenamefont {Fabbris},\ and\ \citenamefont
  {Rosa}}]{Thomas2020}%
  \BibitemOpen
  \bibfield  {author} {\bibinfo {author} {\bibfnamefont {S.~M.}\ \bibnamefont
  {Thomas}}, \bibinfo {author} {\bibfnamefont {F.~B.}\ \bibnamefont {Santos}},
  \bibinfo {author} {\bibfnamefont {M.~H.}\ \bibnamefont {Christensen}},
  \bibinfo {author} {\bibfnamefont {T.}~\bibnamefont {Asaba}}, \bibinfo
  {author} {\bibfnamefont {F.}~\bibnamefont {Ronning}}, \bibinfo {author}
  {\bibfnamefont {J.~D.}\ \bibnamefont {Thompson}}, \bibinfo {author}
  {\bibfnamefont {E.~D.}\ \bibnamefont {Bauer}}, \bibinfo {author}
  {\bibfnamefont {R.~M.}\ \bibnamefont {Fernandes}}, \bibinfo {author}
  {\bibfnamefont {G.}~\bibnamefont {Fabbris}}, \ and\ \bibinfo {author}
  {\bibfnamefont {P.~F.~S.}\ \bibnamefont {Rosa}},\ }\href@noop {} {\bibfield
  {journal} {\bibinfo  {journal} {Sci. Adv.}\ }\textbf {\bibinfo {volume} {6}}
  (\bibinfo {year} {2020})}\BibitemShut {NoStop}%
\bibitem [{\citenamefont {Li}\ \emph {et~al.}(2021)\citenamefont {Li},
  \citenamefont {Nakamura}, \citenamefont {Honda}, \citenamefont {Sato},
  \citenamefont {Homma}, \citenamefont {Shimizu}, \citenamefont {Ishizuka},
  \citenamefont {Yanase}, \citenamefont {Knebel}, \citenamefont {Flouquet},\
  and\ \citenamefont {Aoki}}]{Li2021}%
  \BibitemOpen
  \bibfield  {author} {\bibinfo {author} {\bibfnamefont {D.}~\bibnamefont
  {Li}}, \bibinfo {author} {\bibfnamefont {A.}~\bibnamefont {Nakamura}},
  \bibinfo {author} {\bibfnamefont {F.}~\bibnamefont {Honda}}, \bibinfo
  {author} {\bibfnamefont {Y.~J.}\ \bibnamefont {Sato}}, \bibinfo {author}
  {\bibfnamefont {Y.}~\bibnamefont {Homma}}, \bibinfo {author} {\bibfnamefont
  {Y.}~\bibnamefont {Shimizu}}, \bibinfo {author} {\bibfnamefont
  {J.}~\bibnamefont {Ishizuka}}, \bibinfo {author} {\bibfnamefont
  {Y.}~\bibnamefont {Yanase}}, \bibinfo {author} {\bibfnamefont
  {G.}~\bibnamefont {Knebel}}, \bibinfo {author} {\bibfnamefont
  {J.}~\bibnamefont {Flouquet}}, \ and\ \bibinfo {author} {\bibfnamefont
  {D.}~\bibnamefont {Aoki}},\ }
  {\bibfield  {journal} {\bibinfo  {journal} {Journal of the Physical Society
  of Japan}\ }\textbf {\bibinfo {volume} {90}},\ \bibinfo {pages} {073703}
  (\bibinfo {year} {2021})}\BibitemShut {NoStop}%
\bibitem [{\citenamefont {Hutanu}\ \emph {et~al.}(2020)\citenamefont {Hutanu},
  \citenamefont {Deng}, \citenamefont {Ran}, \citenamefont {Fuhrman},
  \citenamefont {Thoma},\ and\ \citenamefont {Butch}}]{Hutanu2020}%
  \BibitemOpen
  \bibfield  {author} {\bibinfo {author} {\bibfnamefont {V.}~\bibnamefont
  {Hutanu}}, \bibinfo {author} {\bibfnamefont {H.}~\bibnamefont {Deng}},
  \bibinfo {author} {\bibfnamefont {S.}~\bibnamefont {Ran}}, \bibinfo {author}
  {\bibfnamefont {W.~T.}\ \bibnamefont {Fuhrman}}, \bibinfo {author}
  {\bibfnamefont {H.}~\bibnamefont {Thoma}}, \ and\ \bibinfo {author}
  {\bibfnamefont {N.~P.}\ \bibnamefont {Butch}},\ }\href@noop {} {\bibfield
  {journal} {\bibinfo  {journal} {Acta Cryst. B}\ }\textbf {\bibinfo {volume}
  {76}},\ \bibinfo {pages} {137} (\bibinfo {year} {2020})}\BibitemShut
  {NoStop}%
\bibitem [{\citenamefont {Duan}\ \emph {et~al.}(2020)\citenamefont {Duan},
  \citenamefont {Sasmal}, \citenamefont {Maple}, \citenamefont {Podlesnyak},
  \citenamefont {Zhu}, \citenamefont {Si},\ and\ \citenamefont
  {Dai}}]{Duan2020}%
  \BibitemOpen
  \bibfield  {author} {\bibinfo {author} {\bibfnamefont {C.}~\bibnamefont
  {Duan}}, \bibinfo {author} {\bibfnamefont {K.}~\bibnamefont {Sasmal}},
  \bibinfo {author} {\bibfnamefont {M.~B.}\ \bibnamefont {Maple}}, \bibinfo
  {author} {\bibfnamefont {A.}~\bibnamefont {Podlesnyak}}, \bibinfo {author}
  {\bibfnamefont {J.-X.}\ \bibnamefont {Zhu}}, \bibinfo {author} {\bibfnamefont
  {Q.}~\bibnamefont {Si}}, \ and\ \bibinfo {author} {\bibfnamefont
  {P.}~\bibnamefont {Dai}},\ }\href@noop {} {\bibfield  {journal} {\bibinfo
  {journal} {Phys. Rev. Lett.}\ }\textbf {\bibinfo {volume} {125}},\ \bibinfo
  {pages} {237003} (\bibinfo {year} {2020})}\BibitemShut {NoStop}%
\bibitem [{SM()}]{SM}%
  \BibitemOpen
  \href@noop {} {}\bibinfo {note} {See Supplementary Materials for
  details.}\BibitemShut {Stop}%
\bibitem [{Note1()}]{Note1}%
  \BibitemOpen
  \bibinfo {note} {The momentum transfers $\protect \mathbf {Q}=(0,2.1,0)$,
  $(0,1.07,1.07)$, and $(0,1.07,3.07)$ were chosen near, but not exactly at,
  the nuclear Bragg positions $\protect \mathbf {\tau }=(0,2,0)$, $(0,1,1)$,
  and $(0,1,3)$ corresponding to magnetic wavevector $\protect \mathbf {k}=0$
  to avoid contamination by the nuclear Bragg peaks.}\BibitemShut {Stop}%
\bibitem [{\citenamefont {Willa}\ \emph {et~al.}()\citenamefont {Willa} \emph {et~al.},}]{Willa2021}%
  \BibitemOpen
  \bibfield  {author} {\bibinfo {author} {\bibfnamefont {K.}~\bibnamefont
  {Willa} \emph {et~al.}}\ }\href@noop
  {} {\bibinfo  {journal} {to be published}\ }\BibitemShut {NoStop}%
\bibitem [{\citenamefont {Eo}\ \emph {et~al.}()\citenamefont {Eo},
  \citenamefont {Saha}, \citenamefont {Kim}, \citenamefont {Ran}, \citenamefont
  {Horn}, \citenamefont {Hodovanets}, \citenamefont {Collini}, \citenamefont
  {Fuhrman}, \citenamefont {Nevidomskyy}, \citenamefont {Butch}, \citenamefont
  {Fuhrer},\ and\ \citenamefont {Paglione}}]{Eo2021}%
  \BibitemOpen
\bibfield  {journal} {  }\bibfield  {author} {\bibinfo {author} {\bibfnamefont
  {Y.~S.}\ \bibnamefont {Eo}}, \bibinfo {author} {\bibfnamefont {S.~R.}\
  \bibnamefont {Saha}}, \bibinfo {author} {\bibfnamefont {H.}~\bibnamefont
  {Kim}}, \bibinfo {author} {\bibfnamefont {S.}~\bibnamefont {Ran}}, \bibinfo
  {author} {\bibfnamefont {J.~A.}\ \bibnamefont {Horn}}, \bibinfo {author}
  {\bibfnamefont {H.}~\bibnamefont {Hodovanets}}, \bibinfo {author}
  {\bibfnamefont {J.}~\bibnamefont {Collini}}, \bibinfo {author} {\bibfnamefont
  {W.~T.}\ \bibnamefont {Fuhrman}}, \bibinfo {author} {\bibfnamefont {A.~H.}\
  \bibnamefont {Nevidomskyy}}, \bibinfo {author} {\bibfnamefont {N.~P.}\
  \bibnamefont {Butch}}, \bibinfo {author} {\bibfnamefont {M.~S.}\ \bibnamefont
  {Fuhrer}}, \ and\ \bibinfo {author} {\bibfnamefont {J.}~\bibnamefont
  {Paglione}},\ }\href@noop {} {\bibinfo  {journal} {arXiv:2101.03102}\
  }\BibitemShut {NoStop}%
\bibitem [{\citenamefont {Thomas}\ \emph {et~al.}()\citenamefont {Thomas},
  \citenamefont {Stevens}, \citenamefont {Santos}, \citenamefont {Fender},
  \citenamefont {Bauer}, \citenamefont {Ronning}, \citenamefont {Thompson},
  \citenamefont {Huxley},\ and\ \citenamefont {Rosa}}]{Thomas2021}%
  \BibitemOpen
\bibfield  {journal} {  }\bibfield  {author} {\bibinfo {author} {\bibfnamefont
  {S.~M.}\ \bibnamefont {Thomas}}, \bibinfo {author} {\bibfnamefont
  {C.}~\bibnamefont {Stevens}}, \bibinfo {author} {\bibfnamefont {F.~B.}\
  \bibnamefont {Santos}}, \bibinfo {author} {\bibfnamefont {S.~S.}\
  \bibnamefont {Fender}}, \bibinfo {author} {\bibfnamefont {E.~D.}\
  \bibnamefont {Bauer}}, \bibinfo {author} {\bibfnamefont {F.}~\bibnamefont
  {Ronning}}, \bibinfo {author} {\bibfnamefont {J.~D.}\ \bibnamefont
  {Thompson}}, \bibinfo {author} {\bibfnamefont {A.}~\bibnamefont {Huxley}}, \
  and\ \bibinfo {author} {\bibfnamefont {P.~F.~S.}\ \bibnamefont {Rosa}},\
  }\href@noop {} {\bibinfo  {journal} {arXiv:2103.09194}\ }\BibitemShut
  {NoStop}%
\bibitem [{\citenamefont {Niu}\ \emph {et~al.}(2020)\citenamefont {Niu},
  \citenamefont {Knebel}, \citenamefont {Braithwaite}, \citenamefont {Aoki},
  \citenamefont {Lapertot}, \citenamefont {Seyfarth}, \citenamefont {Brison},
  \citenamefont {Flouquet},\ and\ \citenamefont {Pourret}}]{Niu2020}%
  \BibitemOpen
\bibfield  {journal} {  }\bibfield  {author} {\bibinfo {author} {\bibfnamefont
  {Q.}~\bibnamefont {Niu}}, \bibinfo {author} {\bibfnamefont {G.}~\bibnamefont
  {Knebel}}, \bibinfo {author} {\bibfnamefont {D.}~\bibnamefont {Braithwaite}},
  \bibinfo {author} {\bibfnamefont {D.}~\bibnamefont {Aoki}}, \bibinfo {author}
  {\bibfnamefont {G.}~\bibnamefont {Lapertot}}, \bibinfo {author}
  {\bibfnamefont {G.}~\bibnamefont {Seyfarth}}, \bibinfo {author}
  {\bibfnamefont {J.-P.}\ \bibnamefont {Brison}}, \bibinfo {author}
  {\bibfnamefont {J.}~\bibnamefont {Flouquet}}, \ and\ \bibinfo {author}
  {\bibfnamefont {A.}~\bibnamefont {Pourret}},\ }\href@noop {} {\bibfield
  {journal} {\bibinfo  {journal} {Phys. Rev. Lett.}\ }\textbf {\bibinfo
  {volume} {124}},\ \bibinfo {pages} {086601} (\bibinfo {year}
  {2020})}\BibitemShut {NoStop}%
\bibitem [{\citenamefont {Aoki}\ \emph {et~al.}(2013)\citenamefont {Aoki},
  \citenamefont {Knafo},\ and\ \citenamefont {Sheikin}}]{Aoki2013}%
  \BibitemOpen
  \bibfield  {author} {\bibinfo {author} {\bibfnamefont {D.}~\bibnamefont
  {Aoki}}, \bibinfo {author} {\bibfnamefont {W.}~\bibnamefont {Knafo}}, \ and\
  \bibinfo {author} {\bibfnamefont {I.}~\bibnamefont {Sheikin}},\ }\href@noop
  {} {\bibfield  {journal} {\bibinfo  {journal} {C. R. Physique}\ }\textbf
  {\bibinfo {volume} {14}},\ \bibinfo {pages} {53} (\bibinfo {year}
  {2013})}\BibitemShut {NoStop}%
\bibitem [{\citenamefont {Knafo}\ \emph {et~al.}(2009)\citenamefont {Knafo},
  \citenamefont {Raymond}, \citenamefont {Lejay},\ and\ \citenamefont
  {Flouquet}}]{Knafo2009}%
  \BibitemOpen
  \bibfield  {author} {\bibinfo {author} {\bibfnamefont {W.}~\bibnamefont
  {Knafo}}, \bibinfo {author} {\bibfnamefont {S.}~\bibnamefont {Raymond}},
  \bibinfo {author} {\bibfnamefont {P.}~\bibnamefont {Lejay}}, \ and\ \bibinfo
  {author} {\bibfnamefont {J.}~\bibnamefont {Flouquet}},\ }\href@noop {}
  {\bibfield  {journal} {\bibinfo  {journal} {Nat. Phys.}\ }\textbf {\bibinfo
  {volume} {5}},\ \bibinfo {pages} {753} (\bibinfo {year} {2009})}\BibitemShut
  {NoStop}%
\bibitem [{\citenamefont {Haen}\ \emph {et~al.}(1987)\citenamefont {Haen},
  \citenamefont {Flouquet}, \citenamefont {Lapierre}, \citenamefont {Lejay},\
  and\ \citenamefont {Remenyi}}]{Haen1987}%
  \BibitemOpen
  \bibfield  {author} {\bibinfo {author} {\bibfnamefont {P.}~\bibnamefont
  {Haen}}, \bibinfo {author} {\bibfnamefont {J.}~\bibnamefont {Flouquet}},
  \bibinfo {author} {\bibfnamefont {F.}~\bibnamefont {Lapierre}}, \bibinfo
  {author} {\bibfnamefont {P.}~\bibnamefont {Lejay}}, \ and\ \bibinfo {author}
  {\bibfnamefont {G.}~\bibnamefont {Remenyi}},\ }\href@noop {} {\bibfield
  {journal} {\bibinfo  {journal} {J. Low Temp. Phys.}\ }\textbf {\bibinfo
  {volume} {67}},\ \bibinfo {pages} {391} (\bibinfo {year} {1987})}\BibitemShut
  {NoStop}%
\bibitem [{\citenamefont {Paulsen}\ \emph {et~al.}(1990)\citenamefont
  {Paulsen}, \citenamefont {Lacerda}, \citenamefont {Puech}, \citenamefont
  {Haen}, \citenamefont {Lejay}, \citenamefont {Tholence}, \citenamefont
  {Flouquet},\ and\ \citenamefont {de~Visser}}]{Paulsen1990}%
  \BibitemOpen
  \bibfield  {author} {\bibinfo {author} {\bibfnamefont {C.}~\bibnamefont
  {Paulsen}}, \bibinfo {author} {\bibfnamefont {A.}~\bibnamefont {Lacerda}},
  \bibinfo {author} {\bibfnamefont {L.}~\bibnamefont {Puech}}, \bibinfo
  {author} {\bibfnamefont {P.}~\bibnamefont {Haen}}, \bibinfo {author}
  {\bibfnamefont {P.}~\bibnamefont {Lejay}}, \bibinfo {author} {\bibfnamefont
  {J.~L.}\ \bibnamefont {Tholence}}, \bibinfo {author} {\bibfnamefont
  {J.}~\bibnamefont {Flouquet}}, \ and\ \bibinfo {author} {\bibfnamefont
  {A.}~\bibnamefont {de~Visser}},\ }\href@noop {} {\bibfield  {journal}
  {\bibinfo  {journal} {J. Low Temp. Phys.}\ }\textbf {\bibinfo {volume}
  {81}},\ \bibinfo {pages} {317} (\bibinfo {year} {1990})}\BibitemShut
  {NoStop}%
\bibitem [{\citenamefont {Ishida}\ \emph {et~al.}(1998)\citenamefont {Ishida},
  \citenamefont {Kawasaki}, \citenamefont {Kitaoka}, \citenamefont {Asayama},
  \citenamefont {Nakamura},\ and\ \citenamefont {Flouquet}}]{Ishida1998}%
  \BibitemOpen
  \bibfield  {author} {\bibinfo {author} {\bibfnamefont {K.}~\bibnamefont
  {Ishida}}, \bibinfo {author} {\bibfnamefont {Y.}~\bibnamefont {Kawasaki}},
  \bibinfo {author} {\bibfnamefont {Y.}~\bibnamefont {Kitaoka}}, \bibinfo
  {author} {\bibfnamefont {K.}~\bibnamefont {Asayama}}, \bibinfo {author}
  {\bibfnamefont {H.}~\bibnamefont {Nakamura}}, \ and\ \bibinfo {author}
  {\bibfnamefont {J.}~\bibnamefont {Flouquet}},\ }{\bibfield  {journal} {\bibinfo  {journal} {Phys.
  Rev. B}\ }\textbf {\bibinfo {volume} {57}},\ \bibinfo {pages} {R11054}
  (\bibinfo {year} {1998})}\BibitemShut {NoStop}%
\bibitem [{\citenamefont {Ikeda}\ \emph {et~al.}(2006)\citenamefont {Ikeda},
  \citenamefont {Sakai}, \citenamefont {Aoki}, \citenamefont {Homma},
  \citenamefont {Yamamoto}, \citenamefont {Nakamura}, \citenamefont {Shiokawa},
  \citenamefont {Haga},\ and\ \citenamefont {Onuki}}]{Ikeda2006}%
  \BibitemOpen
  \bibfield  {author} {\bibinfo {author} {\bibfnamefont {S.}~\bibnamefont
  {Ikeda}}, \bibinfo {author} {\bibfnamefont {H.}~\bibnamefont {Sakai}},
  \bibinfo {author} {\bibfnamefont {D.}~\bibnamefont {Aoki}}, \bibinfo {author}
  {\bibfnamefont {Y.}~\bibnamefont {Homma}}, \bibinfo {author} {\bibfnamefont
  {E.}~\bibnamefont {Yamamoto}}, \bibinfo {author} {\bibfnamefont
  {A.}~\bibnamefont {Nakamura}}, \bibinfo {author} {\bibfnamefont
  {Y.}~\bibnamefont {Shiokawa}}, \bibinfo {author} {\bibfnamefont
  {Y.}~\bibnamefont {Haga}}, \ and\ \bibinfo {author} {\bibfnamefont
  {Y.}~\bibnamefont {Onuki}},\ }\href@noop {} {\bibfield  {journal} {\bibinfo
  {journal} {J. Phys. Soc. Jpn.}\ }\textbf {\bibinfo {volume} {75}},\ \bibinfo
  {pages} {116} (\bibinfo {year} {2006})}\BibitemShut {NoStop}%
\bibitem [{Note2()}]{Note2}%
  \BibitemOpen
  \bibinfo {note} {The phenomenological inter-ladder magnetic couplings
  $J^b_{inter}$ and $J^c_{inter}$ introduced here may result from two
  successive 'diagonal' inter-ladder exchanges [along third shortest U-U
  distance $d_3$, see Fig. \ref {Fig1}(a-b)].}\BibitemShut {Stop}%
\bibitem [{\citenamefont {Regnault}\ \emph {et~al.}(1999)\citenamefont
  {Regnault}, \citenamefont {Moudden}, \citenamefont {Boucher}, \citenamefont
  {Lorenzo}, \citenamefont {Hiess}, \citenamefont {Vietkin},\ and\
  \citenamefont {Revcolevschi}}]{Regnault1999}%
  \BibitemOpen
  \bibfield  {author} {\bibinfo {author} {\bibfnamefont {L.~P.}\ \bibnamefont
  {Regnault}}, \bibinfo {author} {\bibfnamefont {A.~H.}\ \bibnamefont
  {Moudden}}, \bibinfo {author} {\bibfnamefont {J.~P.}\ \bibnamefont
  {Boucher}}, \bibinfo {author} {\bibfnamefont {E.}~\bibnamefont {Lorenzo}},
  \bibinfo {author} {\bibfnamefont {A.}~\bibnamefont {Hiess}}, \bibinfo
  {author} {\bibfnamefont {A.}~\bibnamefont {Vietkin}}, \ and\ \bibinfo
  {author} {\bibfnamefont {A.}~\bibnamefont {Revcolevschi}},\ }\href@noop {}
  {\bibfield  {journal} {\bibinfo  {journal} {Physica B}\ }\textbf {\bibinfo
  {volume} {259-261}},\ \bibinfo {pages} {1038} (\bibinfo {year}
  {1999})}\BibitemShut {NoStop}%
\bibitem [{\citenamefont {Boullier}(2005)}]{Boullier2005}%
  \BibitemOpen
  \bibfield  {author} {\bibinfo {author} {\bibfnamefont {C.}~\bibnamefont
  {Boullier}},\ }\href@noop {} {\emph {\bibinfo {title} {Ph-D Thesis}}}\
  (\bibinfo  {publisher} {University of Grenoble},\ \bibinfo {year}
  {2005})\BibitemShut {NoStop}%
\bibitem [{\citenamefont {Capogna}\ \emph {et~al.}(2003)\citenamefont
  {Capogna}, \citenamefont {Forgan}, \citenamefont {Hayden}, \citenamefont
  {Wildes}, \citenamefont {Duffy}, \citenamefont {Mackenzie}, \citenamefont
  {Perry}, \citenamefont {Ikeda}, \citenamefont {Maeno},\ and\ \citenamefont
  {Brown}}]{Capogna2003}%
  \BibitemOpen
  \bibfield  {author} {\bibinfo {author} {\bibfnamefont {L.}~\bibnamefont
  {Capogna}}, \bibinfo {author} {\bibfnamefont {E.~M.}\ \bibnamefont {Forgan}},
  \bibinfo {author} {\bibfnamefont {S.~M.}\ \bibnamefont {Hayden}}, \bibinfo
  {author} {\bibfnamefont {A.}~\bibnamefont {Wildes}}, \bibinfo {author}
  {\bibfnamefont {J.~A.}\ \bibnamefont {Duffy}}, \bibinfo {author}
  {\bibfnamefont {A.~P.}\ \bibnamefont {Mackenzie}}, \bibinfo {author}
  {\bibfnamefont {R.~S.}\ \bibnamefont {Perry}}, \bibinfo {author}
  {\bibfnamefont {S.}~\bibnamefont {Ikeda}}, \bibinfo {author} {\bibfnamefont
  {Y.}~\bibnamefont {Maeno}}, \ and\ \bibinfo {author} {\bibfnamefont {S.~P.}\
  \bibnamefont {Brown}},\ }\href@noop {} {\bibfield  {journal} {\bibinfo
  {journal} {Phys. Rev. B}\ }\textbf {\bibinfo {volume} {67}},\ \bibinfo
  {pages} {012504} (\bibinfo {year} {2003})}\BibitemShut {NoStop}%
\bibitem [{\citenamefont {Sato}\ \emph {et~al.}(1988)\citenamefont {Sato},
  \citenamefont {Shamoto}, \citenamefont {Tranquada}, \citenamefont {Shirane},\
  and\ \citenamefont {Keimer}}]{Sato1988}%
  \BibitemOpen
  \bibfield  {author} {\bibinfo {author} {\bibfnamefont {M.}~\bibnamefont
  {Sato}}, \bibinfo {author} {\bibfnamefont {S.}~\bibnamefont {Shamoto}},
  \bibinfo {author} {\bibfnamefont {J.~M.}\ \bibnamefont {Tranquada}}, \bibinfo
  {author} {\bibfnamefont {G.}~\bibnamefont {Shirane}}, \ and\ \bibinfo
  {author} {\bibfnamefont {B.}~\bibnamefont {Keimer}},\ }\href@noop {}
  {\bibfield  {journal} {\bibinfo  {journal} {Phys. Rev. Lett.}\ }\textbf
  {\bibinfo {volume} {61}},\ \bibinfo {pages} {1317} (\bibinfo {year}
  {1988})}\BibitemShut {NoStop}%
\bibitem [{\citenamefont {Pailh\`es}\ \emph {et~al.}(2004)\citenamefont
  {Pailh\`es}, \citenamefont {Sidis}, \citenamefont {Bourges}, \citenamefont
  {Hinkov}, \citenamefont {Ivanov}, \citenamefont {Ulrich}, \citenamefont
  {Regnault},\ and\ \citenamefont {Keimer}}]{Pailhes2004}%
  \BibitemOpen
  \bibfield  {author} {\bibinfo {author} {\bibfnamefont {S.}~\bibnamefont
  {Pailh\`es}}, \bibinfo {author} {\bibfnamefont {Y.}~\bibnamefont {Sidis}},
  \bibinfo {author} {\bibfnamefont {P.}~\bibnamefont {Bourges}}, \bibinfo
  {author} {\bibfnamefont {V.}~\bibnamefont {Hinkov}}, \bibinfo {author}
  {\bibfnamefont {A.}~\bibnamefont {Ivanov}}, \bibinfo {author} {\bibfnamefont
  {C.}~\bibnamefont {Ulrich}}, \bibinfo {author} {\bibfnamefont {L.~P.}\
  \bibnamefont {Regnault}}, \ and\ \bibinfo {author} {\bibfnamefont
  {B.}~\bibnamefont {Keimer}},\ }\href@noop {} {\bibfield  {journal} {\bibinfo
  {journal} {Phys. Rev. Lett.}\ }\textbf {\bibinfo {volume} {93}},\ \bibinfo
  {pages} {167001} (\bibinfo {year} {2004})}\BibitemShut {NoStop}%
\bibitem [{\citenamefont {Xie}\ \emph {et~al.}(2018)\citenamefont {Xie},
  \citenamefont {Wei}, \citenamefont {Gong}, \citenamefont {Fennell},
  \citenamefont {Stuhr}, \citenamefont {Kajimoto}, \citenamefont {Ikeuchi},
  \citenamefont {Li}, \citenamefont {Hu},\ and\ \citenamefont {Luo}}]{Xie2018}%
  \BibitemOpen
  \bibfield  {author} {\bibinfo {author} {\bibfnamefont {T.}~\bibnamefont
  {Xie}}, \bibinfo {author} {\bibfnamefont {Y.}~\bibnamefont {Wei}}, \bibinfo
  {author} {\bibfnamefont {D.}~\bibnamefont {Gong}}, \bibinfo {author}
  {\bibfnamefont {T.}~\bibnamefont {Fennell}}, \bibinfo {author} {\bibfnamefont
  {U.}~\bibnamefont {Stuhr}}, \bibinfo {author} {\bibfnamefont
  {R.}~\bibnamefont {Kajimoto}}, \bibinfo {author} {\bibfnamefont
  {K.}~\bibnamefont {Ikeuchi}}, \bibinfo {author} {\bibfnamefont
  {S.}~\bibnamefont {Li}}, \bibinfo {author} {\bibfnamefont {J.}~\bibnamefont
  {Hu}}, \ and\ \bibinfo {author} {\bibfnamefont {H.}~\bibnamefont {Luo}},\
  }\href@noop {} {\bibfield  {journal} {\bibinfo  {journal} {Phys. Rev. Lett.}\
  }\textbf {\bibinfo {volume} {120}},\ \bibinfo {pages} {267003} (\bibinfo
  {year} {2018})}\BibitemShut {NoStop}%
\bibitem [{\citenamefont {Monthoux}\ and\ \citenamefont
  {Lonzarich}(2001)}]{Monthoux2001}%
  \BibitemOpen
  \bibfield  {author} {\bibinfo {author} {\bibfnamefont {P.}~\bibnamefont
  {Monthoux}}\ and\ \bibinfo {author} {\bibfnamefont {G.~G.}\ \bibnamefont
  {Lonzarich}},\ }\href@noop {} {\bibfield  {journal} {\bibinfo  {journal}
  {Phys. Rev. B}\ }\textbf {\bibinfo {volume} {63}},\ \bibinfo {pages} {054529}
  (\bibinfo {year} {2001})}\BibitemShut {NoStop}%
\bibitem [{\citenamefont {Pustogow}\ \emph {et~al.}(2019)\citenamefont
  {Pustogow}, \citenamefont {Luo}, \citenamefont {Chronister}, \citenamefont
  {Su}, \citenamefont {Sokolov}, \citenamefont {Jerzembeck}, \citenamefont
  {Mackenzie}, \citenamefont {Hicks}, \citenamefont {Kikugawa}, \citenamefont
  {Raghu}, \citenamefont {Bauer},\ and\ \citenamefont {Brown}}]{Pustogow2019}%
  \BibitemOpen
  \bibfield  {author} {\bibinfo {author} {\bibfnamefont {A.}~\bibnamefont
  {Pustogow}}, \bibinfo {author} {\bibfnamefont {Y.}~\bibnamefont {Luo}},
  \bibinfo {author} {\bibfnamefont {A.}~\bibnamefont {Chronister}}, \bibinfo
  {author} {\bibfnamefont {Y.-S.}\ \bibnamefont {Su}}, \bibinfo {author}
  {\bibfnamefont {D.~A.}\ \bibnamefont {Sokolov}}, \bibinfo {author}
  {\bibfnamefont {F.}~\bibnamefont {Jerzembeck}}, \bibinfo {author}
  {\bibfnamefont {A.~P.}\ \bibnamefont {Mackenzie}}, \bibinfo {author}
  {\bibfnamefont {C.~W.}\ \bibnamefont {Hicks}}, \bibinfo {author}
  {\bibfnamefont {N.}~\bibnamefont {Kikugawa}}, \bibinfo {author}
  {\bibfnamefont {S.}~\bibnamefont {Raghu}}, \bibinfo {author} {\bibfnamefont
  {E.~D.}\ \bibnamefont {Bauer}}, \ and\ \bibinfo {author} {\bibfnamefont
  {S.~E.}\ \bibnamefont {Brown}},\ }\href@noop {} {\bibfield  {journal}
  {\bibinfo  {journal} {Nature}\ }\textbf {\bibinfo {volume} {574}},\ \bibinfo
  {pages} {72} (\bibinfo {year} {2019})}\BibitemShut {NoStop}%
\bibitem [{\citenamefont {Steffens}\ \emph {et~al.}(2019)\citenamefont
  {Steffens}, \citenamefont {Sidis}, \citenamefont {Kulda}, \citenamefont
  {Mao}, \citenamefont {Maeno}, \citenamefont {Mazin},\ and\ \citenamefont
  {Braden}}]{Steffens2019}%
  \BibitemOpen
  \bibfield  {author} {\bibinfo {author} {\bibfnamefont {P.}~\bibnamefont
  {Steffens}}, \bibinfo {author} {\bibfnamefont {Y.}~\bibnamefont {Sidis}},
  \bibinfo {author} {\bibfnamefont {J.}~\bibnamefont {Kulda}}, \bibinfo
  {author} {\bibfnamefont {Z.~Q.}\ \bibnamefont {Mao}}, \bibinfo {author}
  {\bibfnamefont {Y.}~\bibnamefont {Maeno}}, \bibinfo {author} {\bibfnamefont
  {I.~I.}\ \bibnamefont {Mazin}}, \ and\ \bibinfo {author} {\bibfnamefont
  {M.}~\bibnamefont {Braden}},\ }\href@noop {} {\bibfield  {journal} {\bibinfo
  {journal} {Phys. Rev. Lett.}\ }\textbf {\bibinfo {volume} {122}},\ \bibinfo
  {pages} {047004} (\bibinfo {year} {2019})}\BibitemShut {NoStop}%
\bibitem [{\citenamefont {Aeppli}\ \emph {et~al.}(1988)\citenamefont {Aeppli},
  \citenamefont {Bucher}, \citenamefont {Broholm}, \citenamefont {Kjems},
  \citenamefont {Baumann},\ and\ \citenamefont {Hufnagl}}]{Aeppli1988}%
  \BibitemOpen
  \bibfield  {author} {\bibinfo {author} {\bibfnamefont {G.}~\bibnamefont
  {Aeppli}}, \bibinfo {author} {\bibfnamefont {E.}~\bibnamefont {Bucher}},
  \bibinfo {author} {\bibfnamefont {C.}~\bibnamefont {Broholm}}, \bibinfo
  {author} {\bibfnamefont {J.~K.}\ \bibnamefont {Kjems}}, \bibinfo {author}
  {\bibfnamefont {J.}~\bibnamefont {Baumann}}, \ and\ \bibinfo {author}
  {\bibfnamefont {J.}~\bibnamefont {Hufnagl}},\ }\href@noop {} {\bibfield
  {journal} {\bibinfo  {journal} {Phys. Rev. Lett.}\ }\textbf {\bibinfo
  {volume} {60}},\ \bibinfo {pages} {615} (\bibinfo {year} {1988})}\BibitemShut
  {NoStop}%
\bibitem [{\citenamefont {Tou}\ \emph {et~al.}(1998)\citenamefont {Tou},
  \citenamefont {Kitaoka}, \citenamefont {Ishida}, \citenamefont {Asayama},
  \citenamefont {Kimura}, \citenamefont {Onuki}, \citenamefont {Yamamoto},
  \citenamefont {Haga},\ and\ \citenamefont {Maezawa}}]{Tou1998}%
  \BibitemOpen
  \bibfield  {author} {\bibinfo {author} {\bibfnamefont {H.}~\bibnamefont
  {Tou}}, \bibinfo {author} {\bibfnamefont {Y.}~\bibnamefont {Kitaoka}},
  \bibinfo {author} {\bibfnamefont {K.}~\bibnamefont {Ishida}}, \bibinfo
  {author} {\bibfnamefont {K.}~\bibnamefont {Asayama}}, \bibinfo {author}
  {\bibfnamefont {N.}~\bibnamefont {Kimura}}, \bibinfo {author} {\bibfnamefont
  {Y.}~\bibnamefont {Onuki}}, \bibinfo {author} {\bibfnamefont
  {E.}~\bibnamefont {Yamamoto}}, \bibinfo {author} {\bibfnamefont
  {Y.}~\bibnamefont {Haga}}, \ and\ \bibinfo {author} {\bibfnamefont
  {K.}~\bibnamefont {Maezawa}},\ }\href@noop {} {\bibfield  {journal} {\bibinfo
   {journal} {Phys. Rev. Lett.}\ }\textbf {\bibinfo {volume} {80}},\ \bibinfo
  {pages} {3129} (\bibinfo {year} {1998})}\BibitemShut {NoStop}%
\bibitem [{\citenamefont {Anderson}(1985)}]{Anderson1985}%
  \BibitemOpen
  \bibfield  {author} {\bibinfo {author} {\bibfnamefont {P.~W.}\ \bibnamefont
  {Anderson}},\ }\href@noop {} {\bibfield  {journal} {\bibinfo  {journal}
  {Phys. Rev. B}\ }\textbf {\bibinfo {volume} {32}},\ \bibinfo {pages} {499}
  (\bibinfo {year} {1985})}\BibitemShut {NoStop}%
\end{thebibliography}

\begin{thebibliography}{4}%
\makeatletter
\providecommand \@ifxundefined [1]{%
 \@ifx{#1\undefined}
}%
\providecommand \@ifnum [1]{%
 \ifnum #1\expandafter \@firstoftwo
 \else \expandafter \@secondoftwo
 \fi
}%
\providecommand \@ifx [1]{%
 \ifx #1\expandafter \@firstoftwo
 \else \expandafter \@secondoftwo
 \fi
}%
\providecommand \natexlab [1]{#1}%
\providecommand \enquote  [1]{``#1''}%
\providecommand \bibnamefont  [1]{#1}%
\providecommand \bibfnamefont [1]{#1}%
\providecommand \citenamefont [1]{#1}%
\providecommand \href@noop [0]{\@secondoftwo}%
\providecommand \href [0]{\begingroup \@sanitize@url \@href}%
\providecommand \@href[1]{\@@startlink{#1}\@@href}%
\providecommand \@@href[1]{\endgroup#1\@@endlink}%
\providecommand \@sanitize@url [0]{\catcode `\\12\catcode `\$12\catcode
  `\&12\catcode `\#12\catcode `\^12\catcode `\_12\catcode `\%12\relax}%
\providecommand \@@startlink[1]{}%
\providecommand \@@endlink[0]{}%
\providecommand \url  [0]{\begingroup\@sanitize@url \@url }%
\providecommand \@url [1]{\endgroup\@href {#1}{\urlprefix }}%
\providecommand \urlprefix  [0]{URL }%
\providecommand \Eprint [0]{\href }%
\providecommand \doibase [0]{http://dx.doi.org/}%
\providecommand \selectlanguage [0]{\@gobble}%
\providecommand \bibinfo  [0]{\@secondoftwo}%
\providecommand \bibfield  [0]{\@secondoftwo}%
\providecommand \translation [1]{[#1]}%
\providecommand \BibitemOpen [0]{}%
\providecommand \bibitemStop [0]{}%
\providecommand \bibitemNoStop [0]{.\EOS\space}%
\providecommand \EOS [0]{\spacefactor3000\relax}%
\providecommand \BibitemShut  [1]{\csname bibitem#1\endcsname}%
\let\auto@bib@innerbib\@empty
\bibitem [{\citenamefont {Brown}(2002)}]{Brown2002}%
  \BibitemOpen
  \bibfield  {author} {\bibinfo {author} {\bibfnamefont {P.~J.}\ \bibnamefont
  {Brown}},\ }\href@noop {} {\emph {\bibinfo {title} {Neutron Data Booklet,
  Chapter 2.5, Magnetic form factors}}}\ (\bibinfo  {publisher} {Institut laue
  Langevin, Grenoble, Edited by A.J. Dianoux and G. Lander},\ \bibinfo {year}
  {2002})\BibitemShut {NoStop}%
\bibitem [{\citenamefont {Fujimori}\ \emph {et~al.}(2021)\citenamefont
  {Fujimori}, \citenamefont {Kawasaki}, \citenamefont {Takeda}, \citenamefont
  {Yamagami}, \citenamefont {Nakamura}, \citenamefont {Homma},\ and\
  \citenamefont {Aoki}}]{Fujimori2021}%
  \BibitemOpen
  \bibfield  {author} {\bibinfo {author} {\bibfnamefont {S.-i.}\ \bibnamefont
  {Fujimori}}, \bibinfo {author} {\bibfnamefont {I.}~\bibnamefont {Kawasaki}},
  \bibinfo {author} {\bibfnamefont {Y.}~\bibnamefont {Takeda}}, \bibinfo
  {author} {\bibfnamefont {H.}~\bibnamefont {Yamagami}}, \bibinfo {author}
  {\bibfnamefont {A.}~\bibnamefont {Nakamura}}, \bibinfo {author}
  {\bibfnamefont {Y.}~\bibnamefont {Homma}}, \ and\ \bibinfo {author}
  {\bibfnamefont {D.}~\bibnamefont {Aoki}},\ }\href@noop {} {\bibfield
  {journal} {\bibinfo  {journal} {J. Phys. Soc. Jpn.}\ }\textbf {\bibinfo
  {volume} {90}},\ \bibinfo {pages} {015002} (\bibinfo {year}
  {2021})}\BibitemShut {NoStop}%
\bibitem [{\citenamefont {Thomas}\ \emph {et~al.}(2020)\citenamefont {Thomas},
  \citenamefont {Santos}, \citenamefont {Christensen}, \citenamefont {Asaba},
  \citenamefont {Ronning}, \citenamefont {Thompson}, \citenamefont {Bauer},
  \citenamefont {Fernandes}, \citenamefont {Fabbris},\ and\ \citenamefont
  {Rosa}}]{Thomas2020}%
  \BibitemOpen
  \bibfield  {author} {\bibinfo {author} {\bibfnamefont {S.~M.}\ \bibnamefont
  {Thomas}}, \bibinfo {author} {\bibfnamefont {F.~B.}\ \bibnamefont {Santos}},
  \bibinfo {author} {\bibfnamefont {M.~H.}\ \bibnamefont {Christensen}},
  \bibinfo {author} {\bibfnamefont {T.}~\bibnamefont {Asaba}}, \bibinfo
  {author} {\bibfnamefont {F.}~\bibnamefont {Ronning}}, \bibinfo {author}
  {\bibfnamefont {J.~D.}\ \bibnamefont {Thompson}}, \bibinfo {author}
  {\bibfnamefont {E.~D.}\ \bibnamefont {Bauer}}, \bibinfo {author}
  {\bibfnamefont {R.~M.}\ \bibnamefont {Fernandes}}, \bibinfo {author}
  {\bibfnamefont {G.}~\bibnamefont {Fabbris}}, \ and\ \bibinfo {author}
  {\bibfnamefont {P.~F.~S.}\ \bibnamefont {Rosa}},\ }\href@noop {} {\bibfield
  {journal} {\bibinfo  {journal} {Sci. Adv.}\ }\textbf {\bibinfo {volume} {6}}
  (\bibinfo {year} {2020})}\BibitemShut {NoStop}%
\bibitem [{\citenamefont {Duan}\ \emph {et~al.}(2020)\citenamefont {Duan},
  \citenamefont {Sasmal}, \citenamefont {Maple}, \citenamefont {Podlesnyak},
  \citenamefont {Zhu}, \citenamefont {Si},\ and\ \citenamefont
  {Dai}}]{Duan2020}%
  \BibitemOpen
  \bibfield  {author} {\bibinfo {author} {\bibfnamefont {C.}~\bibnamefont
  {Duan}}, \bibinfo {author} {\bibfnamefont {K.}~\bibnamefont {Sasmal}},
  \bibinfo {author} {\bibfnamefont {M.~B.}\ \bibnamefont {Maple}}, \bibinfo
  {author} {\bibfnamefont {A.}~\bibnamefont {Podlesnyak}}, \bibinfo {author}
  {\bibfnamefont {J.-X.}\ \bibnamefont {Zhu}}, \bibinfo {author} {\bibfnamefont
  {Q.}~\bibnamefont {Si}}, \ and\ \bibinfo {author} {\bibfnamefont
  {P.}~\bibnamefont {Dai}},\ }\href@noop {} {\bibfield  {journal} {\bibinfo
  {journal} {Phys. Rev. Lett.}\ }\textbf {\bibinfo {volume} {125}},\ \bibinfo
  {pages} {237003} (\bibinfo {year} {2020})}\BibitemShut {NoStop}%
\end{thebibliography}
\end{document}